\documentclass{jpp}
\usepackage{amsmath}
\usepackage{amssymb}
\usepackage{graphicx}
\usepackage[colorlinks,citecolor=blue,linkcolor=blue]{hyperref}
\usepackage{natbib}
\usepackage[utf8]{inputenc}	
\usepackage{nicematrix}
\usepackage{etoolbox}
\makeatletter
\patchcmd{\NAT@citex}
  {\@citea\NAT@hyper@{%
     \NAT@nmfmt{\NAT@nm}%
     \hyper@natlinkbreak{\NAT@aysep\NAT@spacechar}{\@citeb\@extra@b@citeb}%
     \NAT@date}}
  {\@citea\NAT@nmfmt{\NAT@nm}%
   \NAT@aysep\NAT@spacechar\NAT@hyper@{\NAT@date}}{}{}
\patchcmd{\NAT@citex}
  {\@citea\NAT@hyper@{%
     \NAT@nmfmt{\NAT@nm}%
     \hyper@natlinkbreak{\NAT@spacechar\NAT@@open\if*#1*\else#1\NAT@spacechar\fi}%
       {\@citeb\@extra@b@citeb}%
     \NAT@date}}
  {\@citea\NAT@nmfmt{\NAT@nm}%
   \NAT@spacechar\NAT@@open\if*#1*\else#1\NAT@spacechar\fi\NAT@hyper@{\NAT@date}}
  {}{}

\newcommand{\PB}[2]{\left\{ #1 , #2 \right\}}
\renewcommand\Re{\operatorname{Re}}

\shortauthor{A. Hallenbert and G.G. Plunk}

\title{Predicting the Z-pinch Dimits shift through gyrokinetic tertiary instability analysis of the entropy mode}
\author{Axel Hallenbert\aff{1}
  \corresp{\email{axel.hallenbert@ipp.mpg.de}}
 and Gabriel G. Plunk\aff{1}}
 \affiliation{\aff{1}Max-Planck-Institut f\"ur Plasmaphysik, D-17491 Greifswald, Germany}
\date{November 2020}

\begin{document}

\maketitle

\begin{abstract}

    The Dimits shift, an upshift in the onset of turbulence from the linear instability threshold, caused by self-generated zonal flows, can greatly enhance the performance of magnetic confinement plasma devices.  Except in simple cases, using fluid approximations and model magnetic geometries, this phenomenon has proved difficult to understand and quantitatively predict. To bridge the large gap in complexity between simple models and realistic treatment in toroidal magnetic geometries (e.g. tokamaks or stellarators), the present work uses fully gyrokinetic simulations in Z-pinch geometry to investigate the Dimits shift through the lens of tertiary instability analysis, which describes the emergence of drift waves from a zonally dominated state.  Several features of the tertiary instability, previously observed in fluid models, are confirmed to remain. Most significantly, an efficient reduced-mode tertiary model, which previously proved successful in predicting the Dimits shift in a gyrofluid limit (Hallenbert \& Plunk, \textit{J. Plasma Phys}., vol.87, issue 05, 2021, 905870508), is found to be accurate here, with only slight modifications to account for kinetic effects.
    
\end{abstract}

\section{Introduction}\label{SectionIntroduction}

Transport associated with small-scale plasma density/temperature-gradient-driven instabilities \citep{Liewer1985} has proven to be highly significant for nuclear fusion experiments, limiting performance by contributing significant, often stiff, transport \citep{Horton1999,Mantica2009,Ryter2011}. Knowing the associated turbulent transport levels that dictate confinement properties is thus imperative, but an accurate prediction of these is difficult, requiring numerically expensive gyrokinetic simulations \citep{Dimits2000,Lin1998,Parker2004}. 
Therefore, linear instability growth rates, among other things, have been used as a shortcut to expedite this analysis, forming the basis of simpler, more efficient mixing length estimates \citep{Waltz1994,Bourdelle2007a,Pueschel2016}.

However, near marginal linear stability where reactors may be expected to operate due to the aforementioned stiff transport, such a treatment is called into question by the existence of self-generated poloidal zonal flows \citep{Diamond1991,Lin1998,Diamond2005} that efficiently suppress radial streamers by $E\times B$-flow shear decorrelation \citep{Biglari1990a,Lin1998} or zonal-flow-catalysed energy transfer \citep{Terry2018,Terry2021}. In this range this process can be so efficient that transport becomes strongly quenched, resulting in an effective upshift of the critical gradient known as the Dimits shift \citep{Dimits2000}. Unfortunately, though the prediction of the Dimits shift would prove highly valuable, a general quantitative understanding for its exploitation has been lacking. This is because a precise, qualitative description of the upshift has proven elusive, with many competing explanations having been put forth \citep[see e.g.][]{,Ivanov2020,Pueschel2021}. 

The reason why no single explanation has emerged victorious is a result of the veritable wealth of physics present within the Dimits regime, even for the simpler fluid models usually investigated. Though a full overview is beyond the present scope, a few examples are illustrative, the first of which could be categorised as nonlocal transport. \citet{Ivanov2020} observed the presence of radially propagating soliton-like coherent structures, calling them ``ferdinons" due to their resemblance to those previously discovered by \citet{VanWyk2016} in fully developed gyrokinetic turbulence. In addition to these, \citet{Qi2020} also observed transient disordered radial avalanches similar to those of fully developed turbulence \citep{McMillan2009}, which are also known to coexist with such gyrokinetic coherent structures \citep{McMillan2018}. Precisely how these two are related to each other and their effect on the manifestation of a Dimits shift is unknown, but they seem to possess the same root cause while often providing a dominating transport contribution.

Next, although the time-averaged transport by definition is small in the Dimits regime, this does not exclude the possibility of transient turbulent bursts \citep[see][]{Berionni2011}, of which the aforementioned avalanches indeed is an example. Thus, a prevalent feature is predator-prey type energy oscillations between zonal flows, in the role of predator, and drift waves, as prey \citep{Diamond1994}. Now these oscillations are observed to vary qualitatively, both in amplitude, frequency, energy channels, or other ways. This holds even within a single system as the Dimits regime is traversed \citep{Malkov2001,Berionni2011,Kobayashi2015a, Zhu2020}. These ostensibly vanish when continuous finite transport arises around the Dimits threshold, but to what extent there exists a causal link remains unknown. 

Related to the previous features, it is well-known that the strong zonal flows that characterise the Dimits regime cause drift waves and their associated turbulence to localise around those limited regions of zero $E\times B$ flow shear \citep{Kobayashi2012,Kim2018,Kim2019,Zhang2020}. Naturally, the associated transport barriers with limited drift waves in between these points enhances the transport reduction within the Dimits regime. However, an associated nonlinear effect also seems to be that zonal flows eventually saturate into so called $E\times B$-staircases \citep{Dif-Pradalier2010,Peeters2016,Garbet2021} that may extend well above the Dimits regime, raising doubts that the breakdown of these barriers are sufficient to capture the Dimits shift. Nevertheless, though the formation and sustainability of these staircases are poorly understood, they are such a omnipresent feature that they surely must still be accounted for.

Despite the complexity the above examples hint at, attempts continue to be made to encapsulate and predict the Dimits shift. Broadly we can classify these into two categories. The first of these focuses on the turbulent dynamics above the Dimits threshold and attempts to extrapolate downward with the inclusion and emphasis of some particular piece of physics. Naturally, a clear advantage to this approach is how it connects with already existing transport modelling. Unfortunately fully nonlinear kinetic modelling is computationally expensive and it is probable that an accurate prediction of this kind will share this disadvantage. Unfortunately, in the Dimits regime of zonal-flow regulated turbulence, where mean zonal shear exceeds instability growth rates, more efficient conventional quasilinear modelling fails \citep{Pueschel2016,Pueschel2021}. Thankfully, progress continues to be made on this front, with e.q. the combined work of \citet{Pueschel2021} and \citet{Terry2021} predicting a Dimits shift of the appropriate size. This was achieved by extending the turbulence saturation model of \citet{Terry2018} while including the observation of \citet{Hatch2016}, using gyrokinetic pseudospectral analysis, that nonlinear coupling is favoured between unstable modes and their conjugate side band mirror modes, even when these are not present linearly. This is because the three-wave correlation time is inversely proportional to the frequency mismatch, which is minimal for this coupling. Though this approach is promising, it may of course be that some piece of physics not captured by this scheme is in fact be of greater importance, and it remains to be seen which, of the multitude of potential extrapolation candidates, will emerge as most plausible.

The second family of Dimits shift descriptions, to which the present work belongs, instead focuses on the extreme quiescence very close to the linear instability threshold and then proceeds from the linearised behaviour of small amplitude drift waves, in the presence of an essentially static zonal flow, to extrapolate ``upward", i.e. toward the unset of turbulence. Of course this may only prove truly justifiable in the collisionless regime where the zonal flow does not decay collisionally, but it has been observed that the collision rate must be increased considerably to affect the Dimits shift significantly \citep{Weikl2017}. Additionally, however, it is not clear to what extent the more unstable Dimits range remains the same as the more stable range and if the Dimits transition can even be treated as a sharp qualitative transition. Even should this not be the case, investigations of this second kind may nevertheless still be of interest to clarify the Dimits regime, aiding those of the first kind. 

Above all, the main advantage of the second approach is that the zonal flow may be treated as fixed, reducing the problem into an essentially linear one, which should be efficiently solvable. This could prove very useful for reactor optimisation, and in particular stellarators with their large space of possible configurations \citep{Boozer1998}. 
Since it is well known that zonal shear quenching exhibits a complex magnetic geometry dependence \citep{Kinsey2007, Belli2008}, so should the size of the Dimits shift, opening the possibility of linearly unstable configurations that nevertheless are Dimits stable. However, a fruitful search for such configurations require a general Dimits shift prediction.

Some success has been found predicting the Dimits shift for individual cases. However, owing to the complexity of a full gyrokinetic description, these cases all employ some kind of simplified fluid model, predominantly of the Hasegawa-Mima-Wakatani kind \citep{Hasegawa1978,Hasegawa1983}. Frequently, though not always (see e.g. \citet{Ivanov2020}), the analysis underlying these emphasises the so called the tertiary instability \citep{St-Onge2018,Zhu2020,Zhu2020a}. The name of this instability arises from its place in the primary-secondary-tertiary hierarchy \citep{Kim2002,Diamond2005}. Here the primary instability is a linear drift wave instability, then the secondary instability is that which magnifies slight drift wave inhomogeneities to give rise to zonal flows, which finally themselves are subject to the tertiary instability that causes the reemergence of drift waves. When originally studied, this instability, like the secondary instability, was thought to be essentially a plasma nonlinear Kelvin-Helmholtz-type (KH) instability \citep{Rogers2000b}. Subsequent investigations have however revealed this assumption to typically not be true under experimentally relevant conditions. Instead, the dominant tertiary instability is in fact a zonal shear modified primary instability, and so it is vital that linear terms driving the primary instability are included for its treatment \citep{Zhu2020,Zhu2021}.

Though eminently fruitful in elucidating key features, a fluid model cannot encapsulate the full dynamics of full gyrokinetics. Thus a Dimits shift prediction obtained from these models must typically be modified for the kinetic case, which often is not straightforward, and then validated. Owing to the inherent complexity, there has been scant work on this front, which the present article will attempt to begin rectifying. In previous work \citep{Hallenbert2021}, building upon that of \citet{St-Onge2018}, a reduced mode tertiary analysis was employed to predict the Dimits shift of ion temperature gradient (ITG) turbulence in a collisionless gyrofluid limit. Here we proceed to extend that model to the kinetic case, avoiding the question of complex geometries by focusing on the simplest possible one: the Z-pinch.

Previous gyrokinetic simulations of Z-pinch entropy-mode-driven turbulence using the $GS2$-code, while investigating and highlighting various features of near-marginal turbulence, have already established that the Z-pinch indeed exhibits a Dimits shift \citep{Kobayashi2010,Kobayashi2015,Kobayashi2015a}. Here we will again observe similar dynamics in nonlinear simulations using the GENE code \citep{Jenko2000}, but also employ said code to perform sophisticated kinetic tertiary instability analysis and arrive at a Dimits shift prediction to match the nonlinear findings well.

This paper is outlined as follows. The necessary fundamental physics of subsequent analysis is provided in \S\;\ref{SectionBasicPhysics}, with the Z-pinch geometry in \S\;\ref{SectionBasicGeometry}, its collisionless gyrokinetics in \S\;\ref{SectionGyrokinetics}, and some relevant instabilities in \S\;\ref{SectionEntropyMode}-\ref{SectionInterchangeMode}. A description of nonlinear simulations and some key results for this system then follows in \S\;\ref{SectionNonlinearSimulation}, before the tertiary instability is presented in \S\;\ref{SectionTertiaryInstability} and its numerical implementation discussed in \S\;\ref{SectionTertiarySimulations}, the detailed analysis of which is presented in \S\;\ref{SectionSingleMode}-\ref{SectionVelocitySpace}. With this in hand, the reduced mode Dimits prediction is presented, adapted for the present system, and finally tested in \S\;\ref{SectionPrediction}-\ref{SectionComparison}, before a final brief summary and discussion in \S\;\ref{SectionDiscussion}.

\section{Basic Physics}\label{SectionBasicPhysics}

The focus of the present paper will be collisionless gyrokinetics in the Z-pinch geometry. This has the considerable advantage for an exploratory investigation that, when the plasma $\beta$ is low, the dominant electrostatic instability does not exhibit any parallel (along the magnetic field) component, i.e. $k_\parallel=0$ \citep{Simakov2002}, effectively reducing the system from being spatially 3D to just 2D. Additionally zonal flows are then linearly conserved, simplifying and emphasising the tertiary picture of zonal flows since geodesic acoustic modes \citep{Winsor1968,Qiu2018} are not present with their accompanying zonal/drift wave predator-prey type intermittent oscillations  \citep{Miki2007,Kobayashi2015}, and with the Rosenbluth-Hinton remnant fraction \citep{Rosenbluth1998a} being identically one. Thus this system is an excellent test bed when one wishes to extend observations obtained in simplified models \citep[see e.g.][]{Kolesnikov2005,Numata2007,St-Onge2018,Zhu2020} to see how kinetic effects will modify the picture.

\subsection{Z-pinch geometry}\label{SectionBasicGeometry}

The set up for the Z-pinch can be expressed as follows in the usual cylindrical coordinates $(r,\theta,z)$ as follows: a strong current runs in the longitudinal $\hat{\boldsymbol{z}}$-direction which gives rise to a azimuthal magnetic field confining the plasma, assumed to have sufficiently low magnetic $\beta$ so that its presence does not alter the longitudinally uniform magnetic vacuum field
\begin{equation}
    \boldsymbol{B}=B(r)\hat{\boldsymbol{\theta}} \;\;\; \mathrm{where} \;\;\; \frac{1}{B(r)}\frac{dB(r)}{dr}\approx-\frac{1}{r}. 
\end{equation}

Like the magnetic field strength $B$, the large-scale plasma density $n(r)$ and ion/electron temperatures $T_{i}(r)$ and $T_{e}(r)$, which we for simplicity henceforth will assume satisfy $T_i(r)=T_e(r)=T(r)$ since a scaling argument reveals this to be the most unstable scenario \citep{Ricci2006}, only depends on the radial coordinate $r$, since the plasma is free to rapidly equilibrate along the magnetic field lines. In the typical gyrokinetic fashion we are then interested in the small scale fluctuations deriving their energy from the gradients of the large-scale background plasma. To encapsulate these it is advantageous to employ flux tube simulations \citep{Candy2004}, 
and thus we proceed to the local picture, expanding around the point $r=R$, setting $B=B(R)$. Then it is advantageous to introduce the local gradient scale length
\begin{equation}\label{gradients}
    \frac{1}{L_n} = - \left.\frac{1}{n(r)}\frac{dn(r)}{dr}\right|_{r=R}\;\;\; \mathrm{and} \;\;\; \frac{1}{L_T} = - \left.\frac{1}{T(r)}\frac{dT(r)}{dr}\right|_{r=R},
\end{equation}
where $L_n\sim L_r\gg \rho_i$, and adopting a more suitable local flux tube coordinate system $(x',y',z')$ 
    \begin{equation}
        \hat{\boldsymbol{x}}'=\hat{\boldsymbol{r}}, \;\; \hat{\boldsymbol{y}}'=-\hat{\boldsymbol{z}}, \;\; \hat{\boldsymbol{z}}'=\hat{\boldsymbol{\theta}},
    \end{equation}
in which $x'\in(-L_x/2,L_x/2)$ is the radial coordinate, $y'\in(-L_y/2,L_y/2)$ is the bilinear ``poloidal" coordinate and $z$ is the parallel coordinate, where $L_x,L_y\ll R$ are the flux tube box widths. We will however be dropping the prime notation for future convenience and simply write $(x,y,z)$ where it is implicitly understood that we refer to this curvilinear coordinate system and not Cartesian coordinates. Note that, as mentioned previously, parallel variations vanish in the regime of interest and so all quantities will only depend on the perpendicular coordinates $x$ and $y$.

In the absence of flux surfaces, since the magnetic field lines close on themselves after having encircled the central current axis once, the Z-pinch does not, strictly speaking, possess the familiar flux surfaces of toroidal geometries \citep{Diamond2005} that constitute the ``zones" in zonal flows. This is however a minor point, since these can easily be recovered via the inclusion of a small poloidal magnetic field $\delta B\hat{\boldsymbol{y}}$ which twists the individually closed magnetic field line loops into a spiral outlining a ``flux surface cylinder". Taking the limit $\delta B\rightarrow0$ the initial configuration is recovered while this flux surface nevertheless remains.

\subsection{Gyrokinetics}\label{SectionGyrokinetics}

The collisionless gyrokinetic equation \citep{Catto1978,Frieman1982,Abel2013} of ion-scale dynamics for the present system can in Fourier space and dimensionless form be expressed as \citep[see][]{Plunk2014a}
    \begin{equation}\label{gyrokineticequation}
        \left( \frac{\partial}{\partial t} + i \omega_{ds\boldsymbol{k}} \right) g_{s\boldsymbol{k}} + \PB{\Phi_s}{g_s}_{\boldsymbol{k}} = i (\omega_{*s\boldsymbol{k}} - \omega_{ds\boldsymbol{k}}) Z_s \Phi_{s\boldsymbol{k}} f_{0s}
    \end{equation}
for the component with wave number $\boldsymbol{k}$. Here the macroscopic and microscopic spatial dimensions are normalised to $R$ and the ion gyroradius $\rho_i=m_iv_{Ti}/\sqrt{2}eB$ respectively, while the temporal dimension is normalised to the ion streaming time $\sqrt{2}R / v_{Ti}$. The subscript $s=i/e$ denotes the ion/electron species with charge $Z_s e=\pm e$ and macroscopic Maxwellian distribution $f_{0s}(T)$ with mean thermal velocity $v_{Ts} = \sqrt{ 2 T / m_s}$. The corresponding gyrocentre distribution is given by $g_{s\boldsymbol{k}}=J_{0s\boldsymbol{k}}\delta f_{s\boldsymbol{k}}R/\rho_i$, where $\delta f_{s\boldsymbol{k}}$ is the fluctuating distribution, and the Bessel function of the first kind
\begin{equation}
    J_{0s\boldsymbol{k}} = J_{0} (\sqrt{2m_s/m_i} k_\bot v_\bot/v_{Ts})
\end{equation}
encapsulates the Fourier space gyroaverage. The gyroaverage also enters through $\Phi_{s\boldsymbol{k}}=J_{0s\boldsymbol{k}}\varphi_{\boldsymbol{k}}$, where the dimensionless electrostatic potential $\varphi$ is obtained from the electric potential $\phi$ as $\varphi = e \phi R / T \rho_i$. Naturally, the velocity and wavenumber are both split into their parallel and perpendicular components $v_\parallel, k_\parallel, v_\perp, k_\perp$ with respect to the magnetic field.

Next in \eqref{gyrokineticequation}, the Fourier space Poisson bracket, representing the $E\times B$-nonlinearity, is given by
    \begin{equation}\label{PoissonBracket}
        \PB{a}{b}_{\boldsymbol{k}}=\sum_{\boldsymbol{k}_1,\boldsymbol{k}_2}\left(k_{1y}k_{2x}-k_{1x}k_{2y}\right)a_{\boldsymbol{k}_1}b_{\boldsymbol{k}_2}\delta_{\boldsymbol{k},\boldsymbol{k}_1+\boldsymbol{k}_2},
    \end{equation}
where $\delta_{\boldsymbol{k},\boldsymbol{k}_1+\boldsymbol{k}_2}$ is the Kronecker delta. Finally, employing a small magnetic $\beta$ approximation so that $\nabla \mathrm{ln} B = \boldsymbol{b}\cdot\nabla\boldsymbol{b}=\hat{\boldsymbol{x}}/R$, the velocity-dependent diamagnetic and magnetic drift frequencies are given by
    \begin{equation}\label{ConfigurationParameters}
        \omega_{*s\boldsymbol{k}} = -\frac{k_yR}{\sqrt{2}L_n} \left( 1 + \eta \left( \frac{v^2}{v_{Ts}^2}-\frac{3}{2} \right) \right)\;\;\; \mathrm{and} \;\;\;\omega_{ds\boldsymbol{k}} = - \sqrt{2}k_y \left( \frac{v_{\parallel}^2}{v_{Ts}^2} + \frac{ v_{\perp}^2 }{2v_{Ts}^2} \right),
    \end{equation}
where $\eta=L_n/L_T$ and the local gradients $L_n$ and $L_T$ are given by \eqref{gradients}.

To complement the gyrokinetic equation and provide closure, the gyrocentre distribution is related to the potential $\varphi$ via the quasineutrality condition
    \begin{equation}\label{quasineutrality}
        \sum_{s=i,e}Z_s\int d^3v J_{0s\boldsymbol{k}} g_{s\boldsymbol{k}} =  \sum_{s=i,e} n \left( 1 - \Gamma_{0s\boldsymbol{k}} \right) \varphi_{\boldsymbol{k}},
    \end{equation}
where $\Gamma_{0s\boldsymbol{k}}$ can be expressed in terms of the modified Bessel function $I_0$ through
\begin{equation}
    \Gamma_{0s\boldsymbol{k}}=\int d^3vJ_{0s\boldsymbol{k}}^2f_{0s}=I_0\left(\frac{m_sk_\bot^2}{m_i}\right)e^{-\frac{m_sk_\bot^2}{m_i}}.
\end{equation}

\subsection{The entropy mode}\label{SectionEntropyMode}

Before we delve further into the details of the tertiary instability we must be familiar with how the primary linear instability is determined. We thus look for solutions with an $\exp\left(\lambda^p_{\boldsymbol{k}} t\right)$-time dependence, neglecting the nonlinear mode coupling through the Poisson bracket. The partial time derivative $\partial_t$ is symbolically replaced by the primary complex frequency $\lambda^p_{\boldsymbol{k}}=\gamma^p_{\boldsymbol{k}}-i\omega^p_{\boldsymbol{k}}$ in the gyrokinetic equation \eqref{gyrokineticequation}, which is then multiplied by $J_{0s\boldsymbol{k}}$. The resulting equation is then integrated over velocity space and inserted into the quasineutrality condition \eqref{quasineutrality}, resulting in the primary dispersion relation
\begin{equation}\label{PrimaryDispersion}
    D^p(\lambda^p_{\boldsymbol{k}},\boldsymbol{k}) = \sum_{s=i,e}\left( \frac{1}{n} \int d^3v \frac{i (\omega_{*s} - \omega_{ds}) J_{0s\boldsymbol{k}}^2 f_{0s}}{\lambda^p_{\boldsymbol{k}}+i\omega_{ds}} - \left( 1 - \Gamma_{0s\boldsymbol{k}} \right) \right) = 0
\end{equation}
where the $\boldsymbol{k}$-dependence enters through $\omega_{*s}, \omega_{ds}, J_{0s\boldsymbol{k}},$ and $\Gamma_{0s\boldsymbol{k}}$.

Note the full inclusion of kinetic electrons in lieu of a simplified adiabatic response frequently employed for efficiency in gyrokinetic toroidal ITG simulations \citep{Dorland1993} where the Dimits shift has conventionally been studied \citep{Dimits2000}. This is presently necessary because the dominant electrostatic instability at small gradients, originally considered by \citet{Kadomtsev1960}, is the so called entropy mode. This mode, like the MHD interchange mode \citep{Rosenbluth1957}, leaves both the magnetic field and pressure $p=nT$ unchanged, but modifies the entropy $\ln\left(T^{5/2}/p\right)$ via exchange of density $n$ and temperature $T$. Thus it is the plasma analogue of a fluid thermal convective instability enabled by oppositely directed displacements of ions/electrons, and so cannot be encapsulated when electron dynamics are excluded \citep{Ware1962}. 

For the present purpose of investigating the Dimits shift, it is worth noting that the entropy mode is also occasionally referred to as the drift-temperature-gradient mode \citep[see e.g.]{Kesner2000}, despite existing also in the absence of a temperature gradient. This is because the drive term is the same as the ITG mode, and so it is unsurprising that its associated turbulence, subject to the same $E\times B$-nonlinearity, could conceivably be ripe for the same shear-stabilisation characteristic of the Dimits shift, something already confirmed by \citet{Kobayashi2012}. 

Turning our attention back to the dispersion relation \eqref{PrimaryDispersion}, the fact that all linear terms in Equation \eqref{gyrokineticequation} are proportional to $k_y$ makes it clear that the zonal potential
\begin{equation}
    \overline{\varphi}=\frac{1}{L_y}\int_0^{L_y}\varphi dy,
\end{equation}
and thus the zonal $E\times B$-flow $\partial_x\overline{\varphi}$, are completely linearly stable, as previously mentioned, since it is composed of modes with $k_y=0$. Furthermore, as a consequence of us having assumed that $T_e/T_i=1$, it is plain from \eqref{ConfigurationParameters} that the only two remaining independent parameters are $R/L_n$ and $\eta$, which span the configuration space.

As to the solution of \eqref{PrimaryDispersion}, since it is an integral equation that cannot be solved analytically, it must therefore be treated numerically. Though one can, in fact, derive the equivalent entropy mode dispersion relation from a gyrofluid model, a kinetic treatment is necessary close to the marginal stability point of relevance to us, as both \citet{Ricci2006} and \citet{Kobayashi2012} note that the kinetic stability threshold is significantly lower than the collisionless gyrofluid value. Now, instead of explicitly solving \eqref{PrimaryDispersion}, as will be apparent in \S\;\ref{SectionTertiaryInstability}, it is convenient for us to instead initialise a random state and let it evolve according to the gyrokinetic equation \eqref{gyrokineticequation}. The largest growth rate component will then eventually dominate, allowing for easy determination of $\gamma^p$ by fitting an exponential to the system's long-term evolution. 

\begin{figure}
\centering
\includegraphics[width=\linewidth]{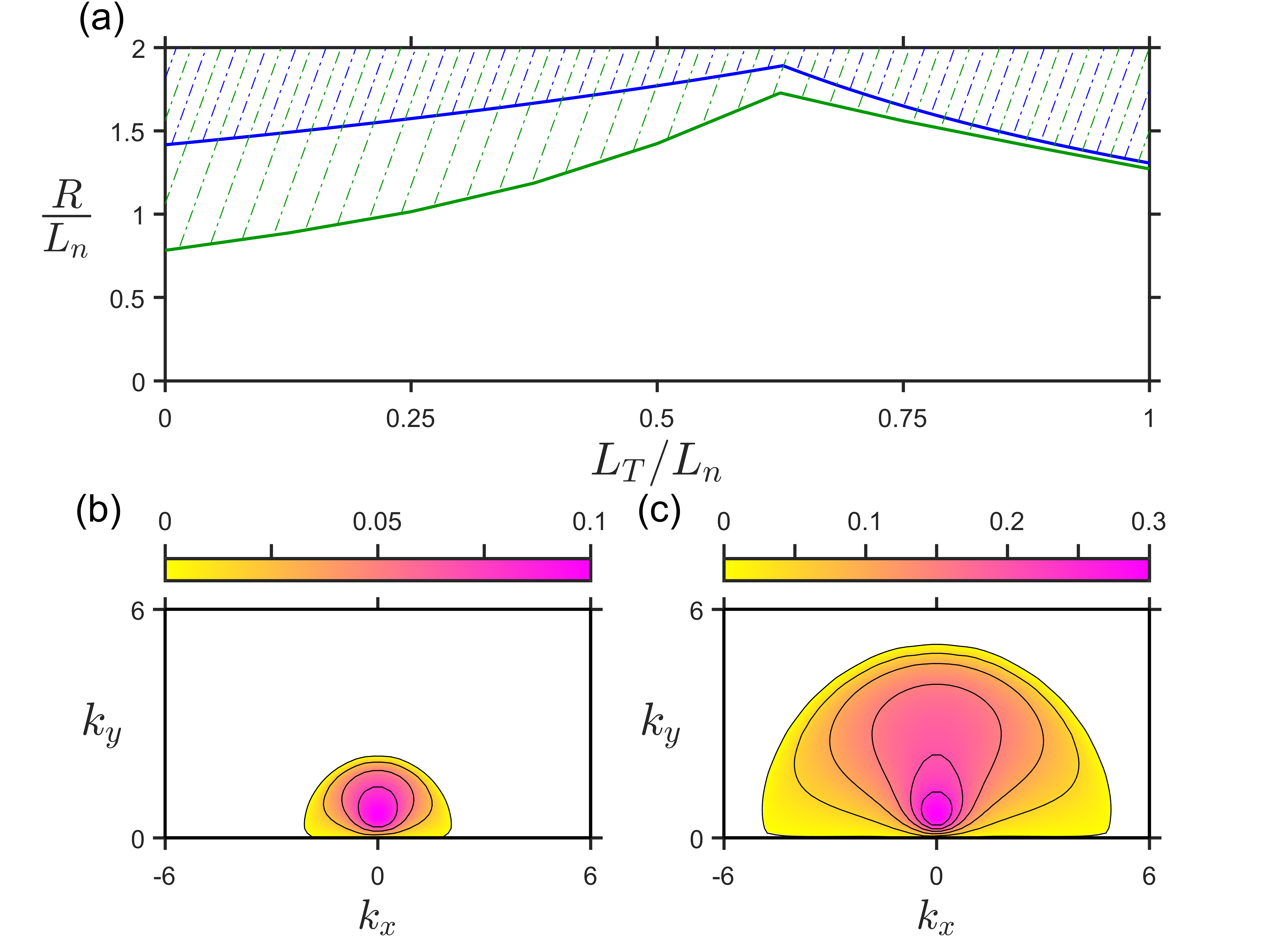}
\caption{(a) Region of collisionless gyrofluid entropy mode instability (blue) and collisionless gyrokinetic entropy mode instability (green). The instability threshold of the latter is seen to occur at significantly lower gradients. In (b) and (c), the entropy mode (primary) growth rate $\gamma^p R/c_s$ as a function of the radial and poloidal wavelengths $k_x,k_y$ is shown for $\eta=0.25$ and (b) $R/L_n=1.4$ and (c) $R/L_n=1.8$, the latter of which corresponds to the Dimits transition. With increasing $R/L_n$ the unstable range is seen to widen significantly, but its maximum remains mostly stationary, peaked around $k_y=0.6$.}
\label{FigureLinear}
\end{figure}

The result of a calculation of this kind, using the code GENE \citep{Jenko2000} with Z-pinch geometry \citep{BanonNavarro2016}, can be seen in figure \ref{FigureLinear}, where (a) shows the range of gyrokinetic instability thus obtained compared to the gyrofluid result of \citet{Ricci2006}. It confirms that the critical gradients are indeed lower than the gyrofluid value, and indicates that $R/L_n$ is a ``good" instability parameters, i.e. one that uniformly destabilises the plasma, unlike $\eta$ or $R/L_T$. In (b) and (c), on the other hand, the entropy mode growth rate is shown as a function of the wavenumber $\boldsymbol{k}$ for $\eta=0.25$ with $R/L_n=1.4$ and $R/L_n=1.8$, which, as we will see in \S\;\ref{SectionDimitsShift}, corresponds to within the Dimits regime and at the Dimits threshold respectively. The result is typical of the Dimits regime, with the mode exhibiting a wide unstable range. However, the growth rate also exhibits a sharp peak around some radial streamer, here $\boldsymbol{k}\approx(0,0.6)$, while other modes constitute a broader, lower growth rate shoulder.

\subsection{The interchange mode}\label{SectionInterchangeMode}

So far we have just been focusing on the gyrokinetic equation and the resulting entropy mode, but the full picture is more complex. At larger gradients there are several MHD instabilities present. Now, the first large scale MHD mode to appear as gradients are increased is the ideal interchange mode \citep{Rosenbluth1957}. Of course we are technically concerned with the completely collisionless case where an MHD treatment is not justified, but the collisionless analogue of this instability can be found by employing the Chew-Goldberger-Low model \citep{Chew1956} and possesses a very similar dispersion relation with a simple instability criterion given by
\begin{equation}\label{InterchangeModeInstability}
    \frac{R}{L_n}\geq \frac{7}{2(1+\eta)}
\end{equation}
when electrons/ions have the same temperature \citep{Ricci2006}.

For our purposes, the presence of the interchange mode at larger gradients, with its characteristic $\boldsymbol{k}$-independent growth rate eliminates the possibility beyond this point of self-generated small-scale and small-amplitude zonal flow stabilisation. Simulations in this range are therefore pathological, exhibiting secular growth at large scales with non-convergent fluxes, and so we dare not attempt to extract any information from these simulations. Still, Z-pinch MHD-like instabilities like this could in principle also be shear-flow stabilised, for example it was predicted that the $m=1$ kink mode \citep{Newcomb1961} could be stabilised \citep{Shumlak1995}, which was also confirmed experimentally \citep{Shumlak2001,Shumlak2009}. These flows are however of a different type, being of significantly larger scales and externally imposed instead of self-generated. As such they affect the background equilibrium, and so we will not be considering the range given by \eqref{InterchangeModeInstability}.

\section{Nonlinear simulations}\label{SectionNonlinearSimulation}

Before proceeding further to discuss the tertiary instability it is suitable to describe how nonlinear simulations are performed, as well as some typical results. As previously mentioned, gyrokinetic simulations were performed using the GENE code \citep{Jenko2000}. Naturally, it is desirable to use as minimal a numerical implementation as possible to allow for the multitude of runs needed to categorise a broad parameter range as belonging to the Dimits regime or not. Thus nonlinear simulations were performed with 32 points in $v_\parallel$ space and 12 points in $\mu$-space. These values were chosen in advance to yield $\gamma^p$-values within a margin of $0.01v_{th}/R$ of the converged value close above the linear instability threshold and and $0.0001v_{th}/R$ around the Dimits threshold. As to the Fourier space decomposition, 128 $k_x$-modes and 16 $k_y$-modes with smallest wavenumber $k_x=k_y=0.1$ was found to prove suitably minimal without compromising key results.

Though collisions were absent in the simulations, explicit numerical dissipation had to be included for stability. However, this dissipation was made not to affect zonal modes in order to eliminate the kind of predator-prey-type dynamics \citep[see e.g.][]{Kobayashi2015a} which would complicate the tertiary picture. Naturally, with the inclusion of any zonal dissipation mechanism (collisional or not) this kind of behaviour would be recovered. Though the net effect on mean transport levels would be negligible for small values, with increasing values this would no longer be the case.

As to the explicit form of the numerical dissipation, dynamically tuned gyroLES hyperviscous dissipation of the form $\nu_x k_x^4+\nu_y k_y^4$ \citep{Morel2011,Morel2012,BanonNavarro2014} was employed. This reduced the excitation of smaller scale modes and simulates the flow of energy to smaller, unresolved scales when turbulence flares up, while flexibly reducing numerical dissipation in the marginal Dimits regime and preventing it from inordinately affecting results. Indeed, this scheme has already been noted as capable of allowing, and being consistent with, a Dimits shift \citep{Morel2011}. Typical values for quiescent, zonally dominated states were $\nu_x\sim10^{-4}$ and $\nu_y\sim10^{-2}$, both of which flared up to $\sim0.05$ in periods of significant turbulence.

\begin{figure}
\centering
\includegraphics[width=0.85\linewidth]{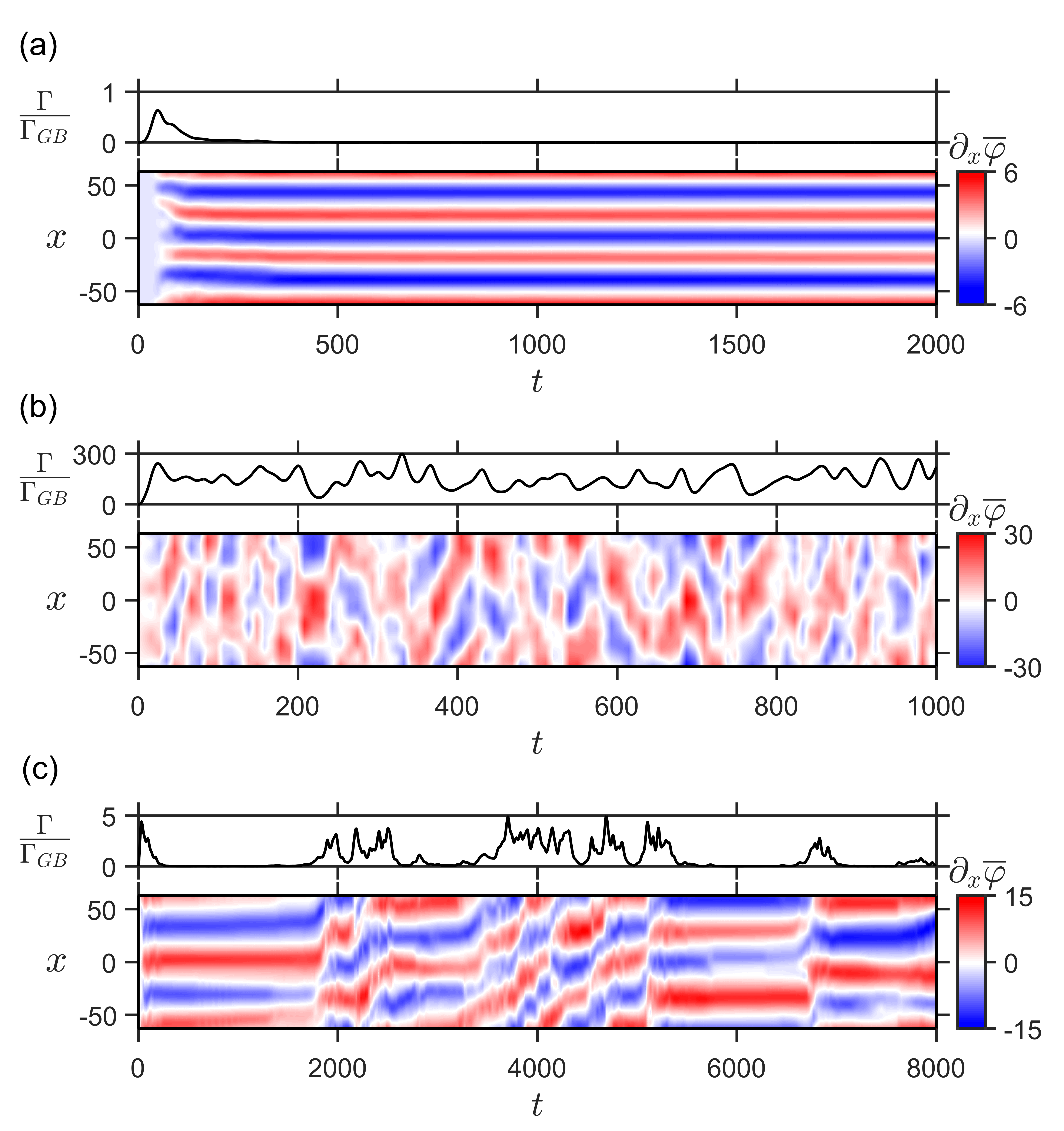}
\caption{Time average particle transport rate $\Gamma$, normalised to the gyro-Bohm level $\Gamma_{GB}$, and the zonal flow $\partial_x\overline{\varphi}$ over time for $\eta=0.25$ and (a) $R/L_n=1.6$, (b) $R/L_n=2.5$, and (c) $R/L_n=2.0$, corresponding to below, well above, and just above the Dimits threshold. In case (a) the potential is virtually unchanged after it initially forms and the transport is very low, while in case (c) the potential is continuously modified and the transport is high. In the intermediary case (b) the system alternates between these two states over extended periods, resulting in bursty transport whose mean magnitude is set by the non-quasistationary state.}
\label{FigureZonalEvolution}
\end{figure}

Some example simulations can be observed in figure \ref{FigureZonalEvolution}, which depicts the box-averaged particle flux
\begin{equation}
    \Gamma = \frac{1}{L_xL_y}\int_0^{L_x}\int_0^{L_y} dxdy \hat{\boldsymbol{x}} \cdot v_{\boldsymbol{E}} \int d^3v g,
\end{equation}
where $v_{\boldsymbol{E}}=\hat{\boldsymbol{z}}\times\nabla\varphi$ is the $E\times B$-drift, as well as the zonal flow $\partial_x\overline{\varphi}$ over time for three different configurations: the first is below the Dimits threshold, the second far above it, and the last is only somewhat above. The dynamics are observed to vary significantly in a way consistent with previous findings \citep[see e.g.][]{Kobayashi2015}. For small $R/L_n$ significant transport only exists transiently before an essentially stable zonal flow is established that effectively quenches drift waves and their associated transport. On the other hand, at large $R/L_n$ the zonal flow continually evolves and fails to quench turbulent transport. At intermediary values, the two types of dynamics coexist at different times, with quasi-stationary zonal flows eventually succumbing in turbulent bursts before being reestablished, a pattern which continually repeats. This is precisely the pattern, dubbed \textit{zonal flow cycling}, observed for the gyrofluid limit described in \citet{Hallenbert2021}, differing significantly only in that the cycling here is much slower.

\subsection{The Dimits shift}\label{SectionDimitsShift}

It has been noted that tokamak turbulence around the Dimits threshold transport exhibit bursts that render time-resolved statistics frustratingly imprecise and possibly misleading \citep{Pueschel2010b}, a feature that has continually plagued investigations of the Dimits shift, in particular when determining whether a given configuration belongs to the Dimits regime or not \citep[for example, compare][]{St-Onge2018,Zhu2020a}. The same holds true here, with statistics averaged over different, similar sections of a burst cycle yielding very different results. Nevertheless, keeping this in mind, employing large enough sample sizes, and focusing on features wherein this problem is lessened, a Dimits threshold can be discerned.

\begin{figure}
\centering
\includegraphics[width=0.85\linewidth]{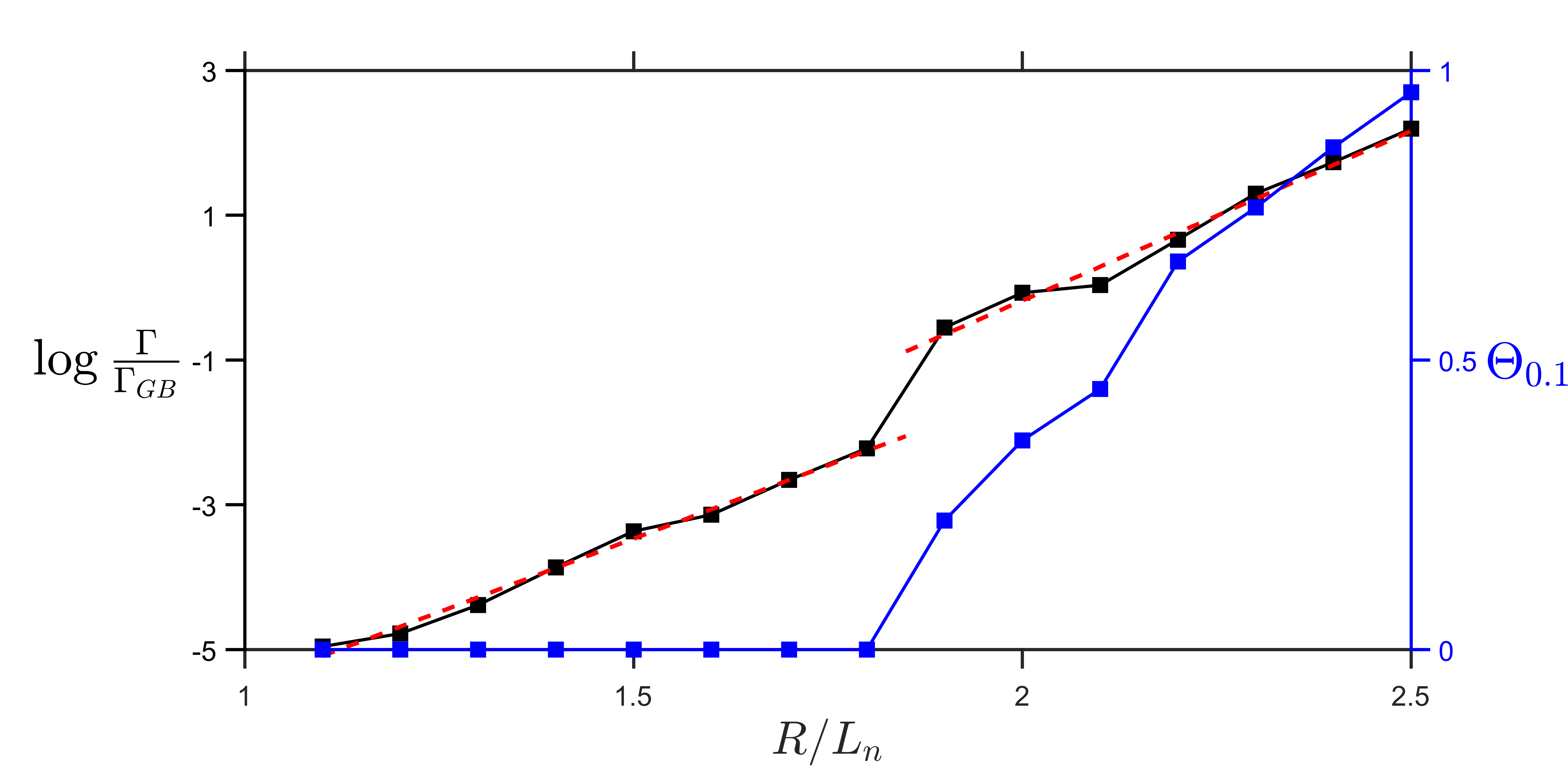}
\caption{In black: particle flux $\Gamma$ as a function of the density gradient $R/L_n$ for $\eta=0.25$. In blue: the fraction $\Theta_{0.1}$ of the simulation time after the initial transport peak where the transport level is higher than 10\% of the initial transport peak, i.e. the fraction of time spent outside of quiescence. The transport is well described by two exponential fits (dashed red) of slopes $4\pm0.3$ and $4.9\pm0.4$, discontinuously jumping between the two at $R/L_n\sim1.8$. This corresponds to that same point at which $\Theta_{0.1}$ becomes greater than 0 and will be identified as the Dimits threshold.}
\label{FigureDimitsTransition}
\end{figure}

Thus we return to what was described in \S\;\ref{SectionNonlinearSimulation}, for which the time-averaged particle flux $\Gamma$ as a function of $R/L_n$ for $\eta=0.25$ can be seen in figure \ref{FigureDimitsTransition}. $\Gamma$ is seen to increase exponentially, but exhibits a discontinuity at $R/L_n\approx1.8$ from a negligible level of $\sim 10^{-2}\Gamma_{GB}$ to the gyro-Bohm value $\sim\Gamma_{GB}$, which we interpret to correspond to the Dimits threshold.

This corresponds to the point where the fraction of time $\Theta$ the system exhibits turbulent bursts, instead of remaining quiescent with quasistatic zonal flows, increases from $0$. To make this quantitative, we introduce
\begin{equation}\label{TurbulentMeasure}
    \Theta_\alpha=\lim_{t\rightarrow\infty}\frac{1}{t}\int_{t_i}^t dt' H\left(\Gamma-\alpha\Gamma(t_i)\right),
\end{equation}
where $H$ is the Heaviside function and $t_i$ is the time when the initial, linear growth phase seizes, i.e. $\Gamma(t_i)$ is taken to be a representative level of turbulent transport. This quantity, for $\alpha=0.1$, is also plotted in figure \ref{FigureDimitsTransition}. It is seen that this fraction begins to increases from $0$ at the Dimits threshold. Though it only gradually increases thereafter, with fully developed turbulence ($\Theta_\alpha\sim 1$) only commencing well above the threshold, this clear signal was used to simply and unambiguously characterise the Dimits threshold.

Now some comments on the observed exponential growth of $\Gamma$ with respect to $R/L_n$ on both sides of the Dimits threshold are in order, since these appear to have differing slopes, $4\pm0.3$ and $4.9\pm0.4$ respectively. Below the Dimits threshold the zonal flow is large and almost stationary, and as we will describe in \S\;\ref{SectionTertiaryInstability}, drift modes will, to lowest order, couple together quasilinearly into marginally (un)stable tertiary modes that only slowly exchange energy with the zonal flow and each other. These are very sensitive to any small perturbation of their dynamics, be it the zonal flow itself, sideband-sideband coupling, or dissipation mechanisms, even of high $k_x$ sidebands. That this is the case has been confirmed by running simulations where hyperviscosity, instead of dynamically changing, was kept constant at comparable values to those obtained through the gyroLES procedure, at which point $\Gamma=0$ was eventually obtained for $R/L_n$-values below $1.2-1.6$, the precise value depending on $\nu_x, \nu_y$. One might be worried that the same holds true for the actual Dimits threshold at $R/L_n\approx1.8$. This is indeed the case, but thankfully this is a much smaller effect, and $\Gamma$ remains quite insensitive at and above the Dimits threshold. To achieve a comparable modification at these values, $\nu_x, \nu_y$ must be increased much further, similar to previous collisional findings where zonal flows are also affected \citep{Weikl2017}.

\begin{figure}
\centering
\includegraphics[width=0.85\linewidth]{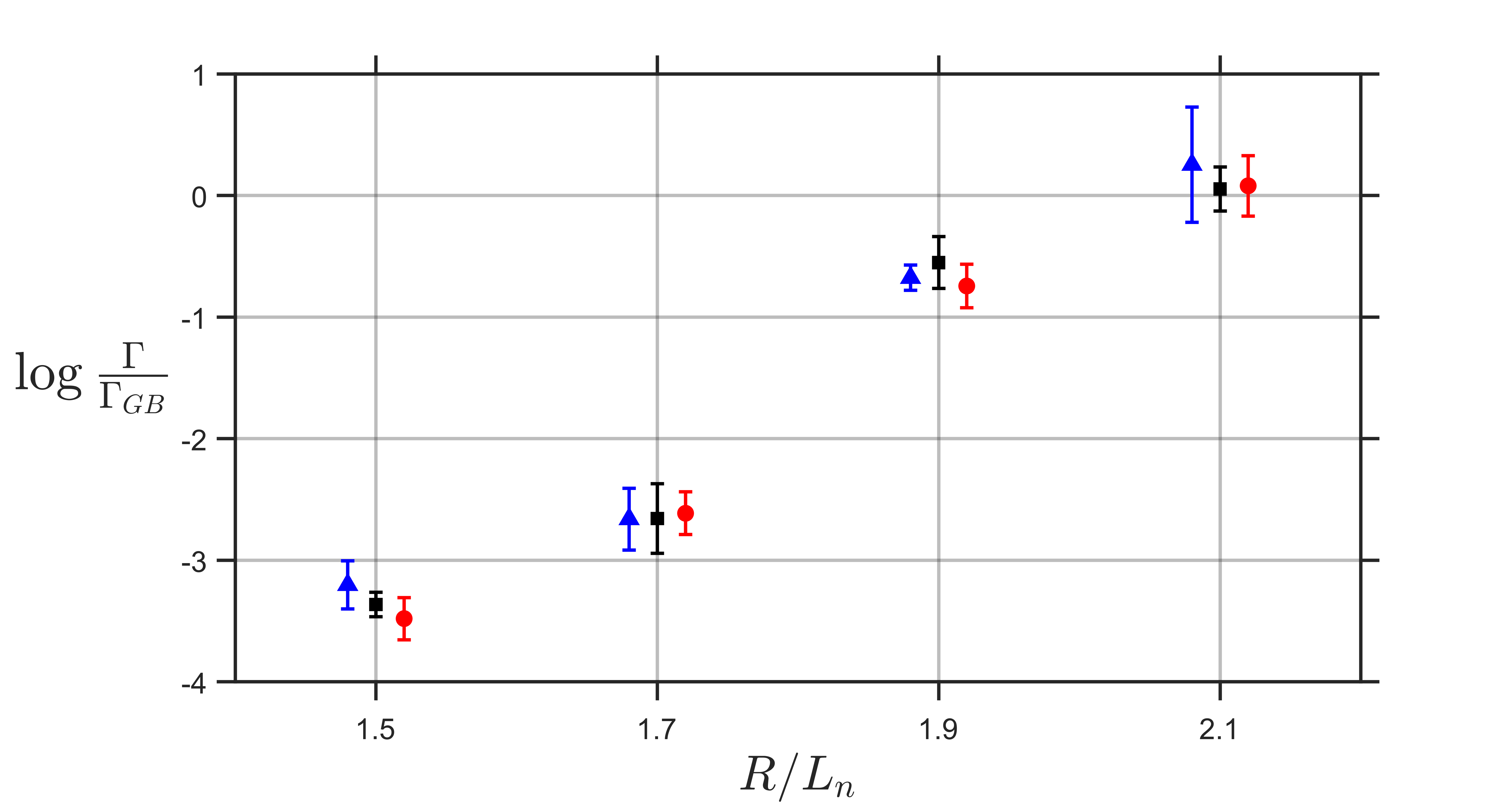}
\caption{Nonlinear particle flux as a function of the density gradient around the observed Dimits threshold $\sim1.8$ for different numerical implementations: standard runs as used in this article (black squares), runs with twice as large box length (blue triangles), and runs with twice the $\boldsymbol{k}$-space resolution (red circles) (the latter two are offset for clarity), with an uncertainty calculated by splitting the full run into ten smaller pieces. Since the observed Dimits transition is seen to be consistent across configurations, the standard configuration employed is seen to be sufficient to determine the Dimits threshold.}
\label{FigureConvergence}
\end{figure}

Turning to the range above the Dimits threshold, the turbulent bursts appear very similar to the continuous turbulence at larger gradients. Unfortunately, this turbulence is not well-behaved, because an unbalanced inverse energy cascade inevitably leads to a pileup, as energy tries to shuffle energy to unresolved large scales, preventing a steady state from being obtained. This feature is well known to plague 2D turbulence \citep{Kraichnan1967,Qian1986,Terry2004}. Thus we stress that, strictly speaking, the turbulent transport levels and their exponential increase above the Dimits threshold should not be trusted. Thankfully, the collisionless Dimits regime is not characterised by turbulence, but rather completely stable or very slowly ``quasilinearly" evolving states, which \textit{are} resolved. The Dimits threshold itself, which occurs when these states fail to remain uniformly quiescent, thus depends on this resolved range. This is confirmed in figure \ref{FigureConvergence} which makes it clear that the discontinuous transport increase at the Dimits threshold remains persistent, and consistent, as the resolution is increased.

\subsection{Zonal shear}\label{SectionZonalShear}

\begin{figure}
\centering
\includegraphics[width=0.85\linewidth]{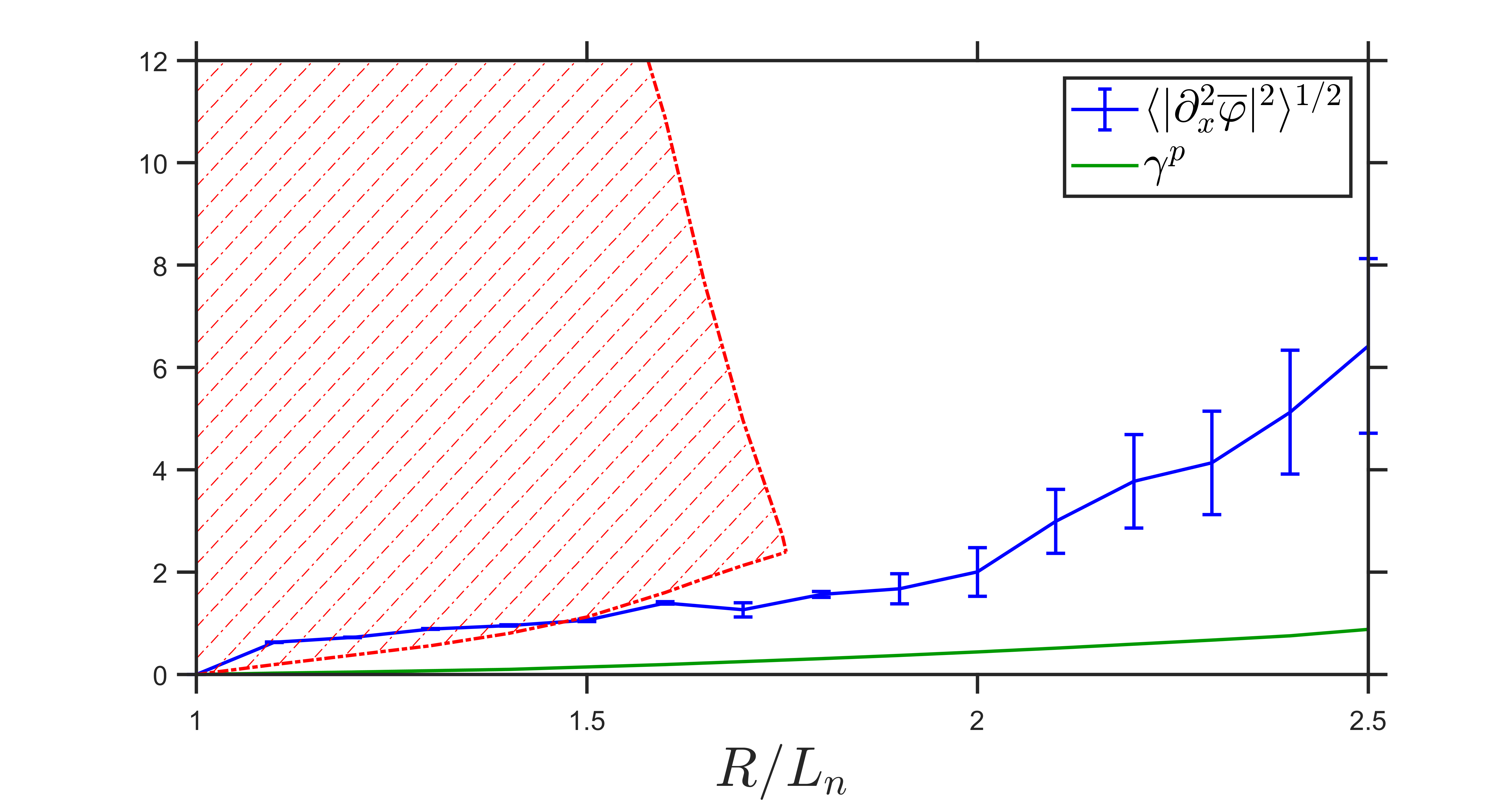}
\caption{Time and box-averaged zonal shear $\langle|\partial_x^2\overline{\varphi}|^2\rangle^{1/2}$, obtained from nonlinear simulations after the transient period, as a function of the density gradient $R/L_n$ for $\eta=0.25$ (blue) compared to the entropy mode growth rate $\gamma^p$ (green), which is significantly lower. Also shown is the range where a sinusoidal zonal profile with $k_x=0.4$, as shown in \S\;\ref{SectionTertiaryInstability} to be approximately maximally stabilising, may be stable (red). It is seen that the nonlinear shear rate remains relatively constant throughout the Dimits regime, at a level around the lower boundary of sinusoidal profile stability, before beginning to gently increase thereafter.}
\label{FigureZonalShear}
\end{figure}

A common quantitative measure of the stabilising effect of zonal flows for turbulence suppression is the box-averaged zonal shear magnitude $\langle|\partial_x^2\overline{\varphi}|^2\rangle^{1/2}$ \citep{Waltz1994,Diamond2005}.  For example, it has been noted that when turbulence is strongly suppressed the quench rule $\langle|\partial_x^2\overline{\varphi}|^2\rangle^{1/2}\sim\gamma^p$ holds \citep{Waltz1998,Kinsey2005}. However, as can be seen in figure \ref{FigureZonalShear}, the nonlinear shear significantly exceeds $\gamma^p$, attaining a value of $\sim 1$ in the Dimits regime. It is likely that this discrepancy can be traced to zonal dissipation, which would push the flow down to the quench rule levels, where turbulence, as found by \citet{Ivanov2020}, acts to reinforce the zonal flow in the Dimits regime. Indeed, \citet{Kobayashi2012} also observed zonal shear levels above the quench rule in the Z-pinch Dimits regime that decreased with increasing collisionality. 

Another heuristic estimate, which forms the basis of simple diffusivity scalings of the kind $1-\alpha\langle|\partial_x^2\overline{\varphi}|^2\rangle^{1/2}/\gamma^p$ with some $\alpha$ that are sometimes employed to model zonal shear quenching \citep{Waltz1998,Kinsey2005}, is that only when $\langle|\partial_x^2\overline{\varphi}|^2\rangle^{1/2}\gtrsim\gamma^p$ will zonal flows have a significant effect on turbulent saturation \citep{Xanthopoulos2007,Pueschel2010b}. As is clear in figure \ref{FigureZonalShear}, this presently holds, but $\langle|\partial_x^2\overline{\varphi}|^2\rangle^{1/2}$ increases faster than $\gamma^p$ above the Dimits threshold, even when accounting for the fact that the effective shearing decorrelation rate decreases when the zonal flow varies rapidly in time, like described by \citet{Hahm1999}. A similar increase in ITG simulations was noted by \citet{Terry2021} as being inconsistent with the tertiary picture of the Dimits shift, seemingly under the assumption that greater zonal shear should be more stabilising. However, that this need not be the case has already been observed \citep[see e.g.][]{Kinsey2005, Kobayashi2012}, and by also plotting the approximate range zonal shear amplitudes (determined as outlined in \S\;\ref{SectionTertiaryInstability} and \S\;\ref{SectionPrediction}) where a sinusoidal profile can be completely stable in figure \ref{FigureZonalShear}, it is seen to eventually be reduced to a single value, both from above and below, close to the Dimits threshold.

\subsection{Localisation}\label{SectionLocalisation}

\begin{figure}
\centering
\includegraphics[width=0.95\linewidth]{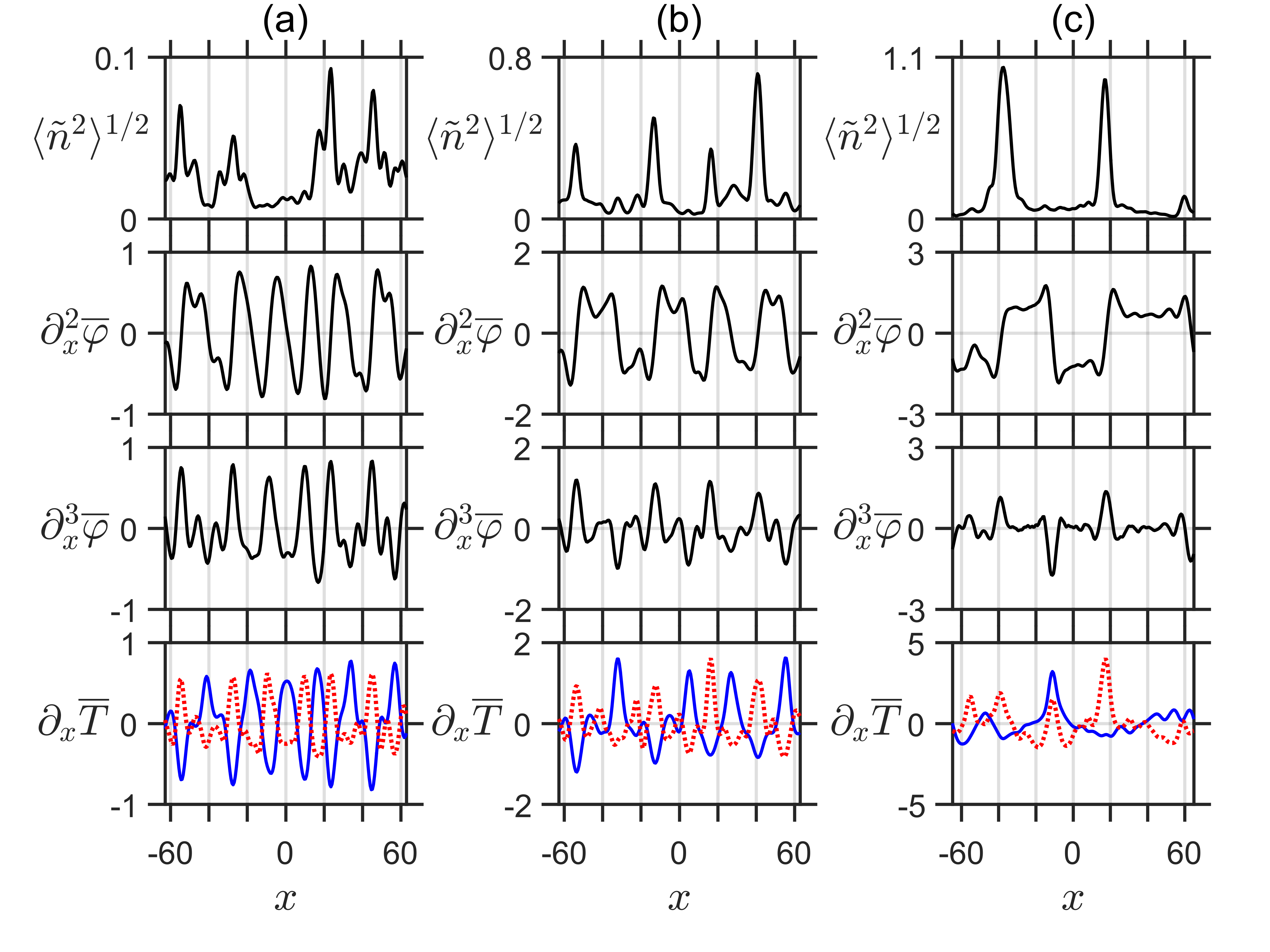}
\caption{Mean drift wave density perturbation $\tilde{n}$ as a function of the radial coordinate $x$ in the presence of the quasistationary zonal flow $\partial_x\overline{\varphi}$ and zonal fluctuating temperature $\overline{T}$ for: (a) $R/L_n=1.2$, (a) $R/L_n=1.5$, and (c) $R/L_n=1.8$. As can be seen, the ion temperature (blue) and the electron temperature (red, dotted) tend to display opposite signs over a majority of the region. Meanwhile, it can be seen that as the gradient is increased towards the Dimits threshold the stable zonal flow by necessity increasingly resembles a staircase state with broadening `steps' of nearly constant zonal shear $\partial^2_x\overline{\varphi}$. Meanwhile, the drift waves increasingly localise around zonal flow minima, i.e. where the zonal shear  $\partial^2_x\overline{\varphi}$ vanishes and $\partial^3_x\overline{\varphi}>0$.}
\label{FigureTertiaryLocalisation}
\end{figure}

A persistent feature of marginal drift waves in the presence of a zonal flow is the localisation of said modes to regions of zero zonal shear, $\partial^2_x\overline{\varphi}=0$. This is in accordance with the intuitive picture that the zonal flow at these points is unable to decorrelate radial streamers, effectively only transporting them poloidally, i.e. in the $y$-direction, in a uniform way \citep{Ivanov2020}. 
However, this necessary condition is not sufficient for drift waves to congregate at such a point, since they asymmetrically respond to minima $\partial_x^3\overline{\varphi}>0$ and maxima $\partial_x^3\overline{\varphi}<0$: significant localisation is only observed at minima \citep{McMillan2011,Zhu2020}. That the same holds here is very clear in figure \ref{FigureTertiaryLocalisation}, where it is additionally observed that more developed staircase states, i.e. broader rectangular shapes \citep{Dif-Pradalier2010,Peeters2016} of nearly constant shear interspersed with sharp transitions from positive to negative values, arises as one increases $R/L_n$ to approach the Dimits threshold from below. 

In a modified Hasegawa-Mima system, the formation of well-defined staircase states like these was recently conjectured by \citet{Zhang2020} to provide effective transport barriers of drift waves, and in particular drifting coherent structures like ferdinons that carry transport \citep{Ivanov2020}, reflecting and trapping them in the vicinity of zonal extrema. This effect was even boldly speculated to be more important than zonal shearing for transport quenching. Though possibly relevant, particularly close to marginal stability, this conclusion seems dubious for the system at hand. This is because drifting structures were not observed. Instead drift waves, though indeed localised, commonly remained at \textit{very low} amplitudes with essentially no self-interactions for extended periods of time, seemingly unable to grow in the presence of strong shear.

As a final note before we proceed, it is apparent from figure \ref{FigureTertiaryLocalisation} that ion/electron temperature moments predominantly self-organise in such a way that they are out of phase with each other, something we will clarify further in \S\;\ref{SectionElectronIonPhase}-\ref{SectionVelocitySpace}.

\section{The tertiary instability}\label{SectionTertiaryInstability}

Having described how nonlinear simulations were performed and some typical results, we now consider a static, high amplitude zonal profile $\overline{\varphi}$ and proceed like in \S\;\ref{SectionEntropyMode} to study the tertiary instability in greater detail, since it served as the motivation for the present study. Naturally we can no longer neglect the nonlinear term, but, as is plain from \eqref{PoissonBracket}, the $E\times B$ nonlinearity $\PB{\Phi_s}{g_s}_{\boldsymbol{k}}$ only couples modes together if they satisfy the triplet condition 
\begin{equation}\label{TripletCondition}
\boldsymbol{k}=\boldsymbol{k}'+\boldsymbol{k}''.
\end{equation}
Thus, zonal modes with $\boldsymbol{k}=(k_x,0)$ only couple modes sharing the same poloidal wavenumber, and for small amplitude drift waves the nonlinear self-interaction between modes of different $k_y$ can be neglected. Because of this, the tertiary instability problem is characterised by a single poloidal wavenumber, and the discrete spectral decomposition that the vector wavenumbers given by
\begin{equation}\label{BeforeGalerkin}
    \boldsymbol{p}, \boldsymbol{p}\pm\boldsymbol{q}, \boldsymbol{p}\pm2\boldsymbol{q} ... \; ,
\end{equation}
where $\boldsymbol{p}=(0,p)$ is purely poloidal and $\boldsymbol{q}=(q,0)$ purely radial, should be considered to constitute a combined tertiary mode
\begin{equation}
    e^{\gamma^t t+ipy}\sum_{s,n} a_{sn}e^{inqx}g_{s\boldsymbol{p}+n\boldsymbol{q}}
\end{equation}
where $n$ runs over the integers and $a_n$ have to be determined by insertion into the gyrokinetic equation. 

Galerkin-truncating \eqref{BeforeGalerkin} after some $\boldsymbol{q}_\mathrm{G}=j\boldsymbol{q}$ we thus have a finite set of modes whose radial wavenumbers can be combined into a $(2j+1)$-dimensional vector
\begin{equation}\label{CapitalQ}
    \mathsfbi{Q} = \begin{bmatrix}-q_\mathrm{G}\\q_\mathrm{G}+q\\...\\q_\mathrm{G}-q\\q_\mathrm{G}\end{bmatrix}
\end{equation}
whose elements we denote by $Q_m$, where $-j\leq m\leq j$. Then the Fourier components of other variables can also be combined into the corresponding vectors
\begin{align}
    \mathsfbi{\tilde{g}}_s=&\begin{bmatrix}g_{s\boldsymbol{p}-\boldsymbol{q}_\mathrm{G}}\\g_{s\boldsymbol{p}-\boldsymbol{q}_\mathrm{G}+\boldsymbol{q}}\\...\\g_{s\boldsymbol{p}+\boldsymbol{q}_\mathrm{G}-\boldsymbol{q}}\\g_{s\boldsymbol{p}+\boldsymbol{q}_\mathrm{G}}\end{bmatrix}, 
    \;\;\;\; 
    \mathsfbi{\overline{g}}_s=\begin{bmatrix}g_{s-\boldsymbol{q}_\mathrm{G}}\\g_{s-\boldsymbol{q}_\mathrm{G}+\boldsymbol{q}}\\...\\g_{s\boldsymbol{q}_\mathrm{G}-\boldsymbol{q}}\\g_{s\boldsymbol{q}_\mathrm{G}}\end{bmatrix}, 
    \;\;\;\; \nonumber\\
    \boldsymbol{\tilde{\varphi}}=&\begin{bmatrix}\varphi_{\boldsymbol{p}-\boldsymbol{q}_\mathrm{G}}\\\varphi_{\boldsymbol{p}-\boldsymbol{q}_\mathrm{G}+\boldsymbol{q}}\\...\\\varphi_{\boldsymbol{p}+\boldsymbol{q}_\mathrm{G}-\boldsymbol{q}}\\\varphi_{\boldsymbol{p}+\boldsymbol{q}_\mathrm{G}}\end{bmatrix}, 
    \;\;\;\;\;\;
    \boldsymbol{\overline{\varphi}}=\begin{bmatrix}\varphi_{-\boldsymbol{q}_\mathrm{G}}\\\varphi_{-\boldsymbol{q}_\mathrm{G}+\boldsymbol{q}}\\...\\\varphi_{\boldsymbol{q}_\mathrm{G}-\boldsymbol{q}}\\\varphi_{\boldsymbol{q}_\mathrm{G}}\end{bmatrix},
\end{align}
with elements $\tilde{g}_{sm}, \overline{g}_{sm}, \tilde{\varphi}_{sm}$, and $\overline{\varphi}_{sm}$, for the potential and gyrocentre distributions. Letting $\mathsfbi{J}_{0s}$, and $\boldsymbol{\Gamma}_{0s}$ denote the corresponding diagonal matrices, i.e. with $m$th entries $J_{0smm}=J_0(\sqrt{2(p^2+Q_m^2)m_s/m_i}w_s)$ and $\Gamma_{0smm}=I_0((p^2+Q_m^2)m_s/m_i)e^{-(p^2+Q_m^2)m_s/m_i}$, we can thus write the set of coupled equations given by \eqref{gyrokineticequation} and \eqref{quasineutrality} for these modes as
\begin{equation}\label{gyrokineticsystem}
    \mathsfbi{A}_s\mathsfbi{\tilde{g}}_s = \mathsfbi{B}_s\mathsfbi{J}_{0s}\boldsymbol{\tilde{\varphi}}
\end{equation}
and
\begin{equation}\label{quasineutralitysystem}
        \sum_{s=i,e}Z_s\int d^3v \mathsfbi{J}_{0s} \mathsfbi{\tilde{g}}_s =  \sum_{s=i,e} n \left( \mathsfbi{I} - \boldsymbol{\Gamma}_{0s} \right) \boldsymbol{\tilde{\varphi}},
    \end{equation}
where $\mathsfbi{I}$ is the identity matrix and the matrices $\boldsymbol{A}_s$ and $\boldsymbol{B}_s$ are given by
\begin{equation}
\boldsymbol{A}_s=(\lambda+i\omega_{ds\boldsymbol{p}})\mathsfbi{I}+\mathsfbi{C}_{1s}, \;\;\; \boldsymbol{B}_s=iZ_sf_{0s}(\omega_{*s\boldsymbol{p}}-\omega_{ds\boldsymbol{p}})\mathsfbi{I}+\mathsfbi{C}_{2s},
\end{equation}
while the Poisson bracket is encapsulated by the matrices $\mathsfbi{C}_{1s}$ and $\mathsfbi{C}_{2s}$ with elements
\begin{equation}
    C_{1smn}=-p\sum_{l=-j}^{j}Q_lJ_{0sll}\overline{\varphi}_l\delta_{m(l+n)}, \;\;\; C_{2smn}=-p\sum_{l=-j}^{j}Q_l\overline{g}_{sl}\delta_{m(l+n)}.
\end{equation}

The procedure is now analogous to that of the primary instability, so upon multiplying \eqref{gyrokineticsystem} by $\mathsfbi{A}^{-1}$ and substituting the result into \eqref{quasineutralitysystem} we find that, in order for the solution to possess nonzero $\boldsymbol{\tilde{\varphi}}$, the tertiary dispersion relation
\begin{equation}\label{tertiaryCompleteDispersion}
    \det\left[\sum_{s=i,e}\left(\frac{Z_s}{n}\int d^3v\mathsfbi{J}_{0s}\mathsfbi{A}_s^{-1}\mathsfbi{B}_s\mathsfbi{J}_{0s}-(\mathsfbi{I}-\boldsymbol{\Gamma}_{0s})\right)\right]=0
\end{equation}
must be satisfied. The exact solution can then recovered in the continuum limit where $q_\mathrm{G}\rightarrow\infty$, though we will only be concerned with the Galerkin-truncated system.

The similarity of \eqref{PrimaryDispersion} and \eqref{tertiaryCompleteDispersion} as presented here is deliberate: we consider the tertiary instability to precisely be a linear confluence of the poloidal band of primary modes coupled together by the zonal flow. Though this allows the description of a KH-like mode, like the tertiary mode is frequently thought to be \citep[see e.g.][]{Kolesnikov2005a,Numata2007}, it also allows modes of an entirely different kind. This is the modified primary instability, which differs in that the energy feeding the instability is supplied by the background gradients instead of the zonal flow, a feature elucidated and emphasised by \citet{Zhu2020}.

\subsection{Tertiary instability simulations}\label{SectionTertiarySimulations}

Of course, in general \eqref{tertiaryCompleteDispersion}, when expanded out, becomes an intractable sum of products of integrals like \eqref{PrimaryDispersion} but with ever more complex integrands. Since \eqref{PrimaryDispersion} already must be solved numerically, the same holds true here, but as $q_\mathrm{G}$ is increased it becomes cumbersome to do so directly. Instead the alternate, aforementioned procedure of evolving the gyrokinetic equation directly is employed, in a procedure that is essentially as described in \S\;\ref{SectionEntropyMode}. 

A zonal profile, i.e. the set of $2n$ zonal distributions $\overline{g}_{s\boldsymbol{k}}$ with potential modes $\overline{\varphi}_{\boldsymbol{k}}$, is either extracted from nonlinear simulations or initialised from scratch into some desired configuration. Then a single poloidal wavenumber $p$ is chosen and a new nonlinear simulation with an $n$ $k_x$-mode and $2$ $k_y$-mode ($k_y=0$ and $k_y=p$) Fourier decomposition is initialised with this zonal profile and small amplitude drift waves. The system is then allowed to evolve nonlinearly, but with the zonal profile frozen in place. By the triplet condition \eqref{TripletCondition} drift wave self-interaction are then not present, and having frozen the zonal profile the drift wave ``back reaction" is also removed. Combined, this enables the convergence of the observed tertiary growth rate without drift waves nonlinearly affecting the zonal profile or each other as they grow in amplitude

Before proceeding it is worth discussing the matter of numerical damping and resolution for this procedure, since this was found to be important. Performing tertiary simulations of stable profiles extracted from low $R/L_n$-simulations, these were observed to only be marginally stable. Changing $\nu_x$ by a small amount, though it might be expected to have little effect, therefore nevertheless often causes the profile to become unstable. Thus it seems that nonlinearly obtained profiles are ``balanced" for tertiary sideband ``resonant" coupling that is finely modulated by $\nu_x$. This might make one nervous that this numerical dissipation has a nonphysically important role when establishing stability. However, this is not the case. Modifying $\nu_x$ to destabilise an initially stable profile, after restarting a nonlinear simulation it will after some transient period of increased drift waves quickly be only slightly modified to restore tertiary stability. Therefore, the specific realisation of $\nu_x$, unless massively increased, does not matter for the formation of stable profiles. The situation is entirely analogous for the numerical resolution, where the removal or inclusion of further modes can destabilise specific instances of zonal profiles, but not change whether stability is expected to nonlinearly arise so long as the resolution is not taken to be patently too low.

\begin{figure}
\centering
\includegraphics[width=0.85\linewidth]{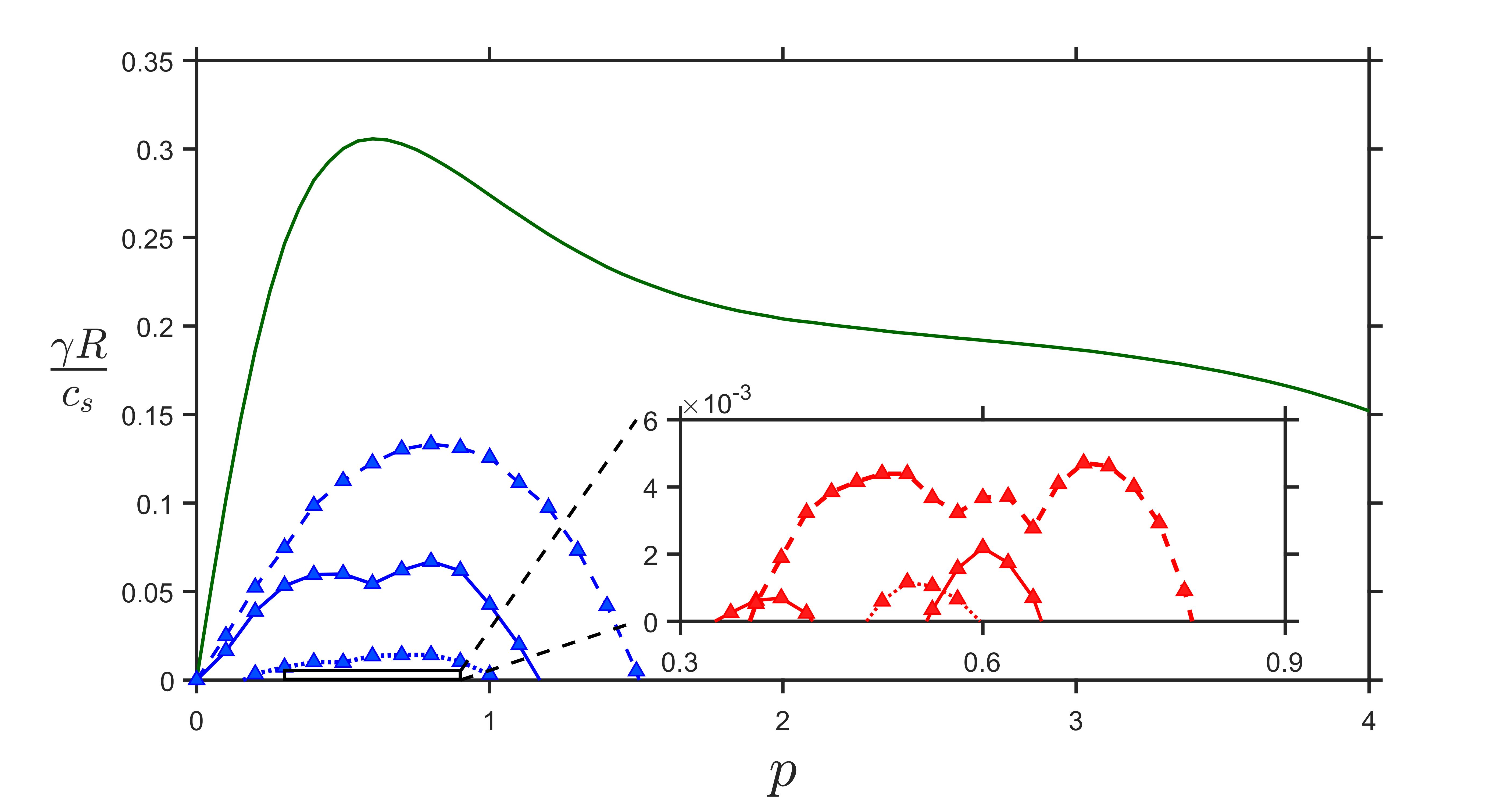}
\caption{Tertiary growth rates of different, randomly chosen zonal profiles, obtained from nonlinear simulations with $\eta=0.25$ and $R/L_n=1.8$, as a function of the poloidal wave number $p$. Blue lines correspond to zonal profiles taken during turbulent periods, while red (inlaid) lines correspond to profiles taken during quiescent periods. Additionally, the primary growth rate $\gamma^\mathrm{p}$ is plotted in green for comparison. As can be seen, the quiescent profiles are only unstable to a few $p$-values clustered around $\sim$0.6, the most primary unstable point, with growth rates severely reduced to about 1\% of the primary growth rates. In comparison, the turbulent profiles are much more unstable to a broader range of $p$-values.}
\label{FigureTertiaryKy}
\end{figure}

To begin investigating the present tertiary instability using the aforementioned scheme, in figure \ref{FigureTertiaryKy} the tertiary growth rate $\gamma^t$ of the poloidal band $p$ under the influence of different zonal profiles extracted from nonlinear simulations are are compared with the primary growth rate $\gamma^p$. The profiles are obtained at the Dimits threshold $\eta=0.25$ and $R/L_n=1.8$, where turbulent bursts first manifest. It is seen that $\gamma^t$ is much greater for zonal profiles of turbulent periods compared with their quasistationary counterparts that are only barely unstable with $\gamma^t\sim0.01\gamma^p$. However in both cases the growth rate is seen to peak around the same value $p\approx0.6$ where $\gamma^p$ attains its maximum.

\begin{figure}
\centering
\includegraphics[width=0.85\linewidth]{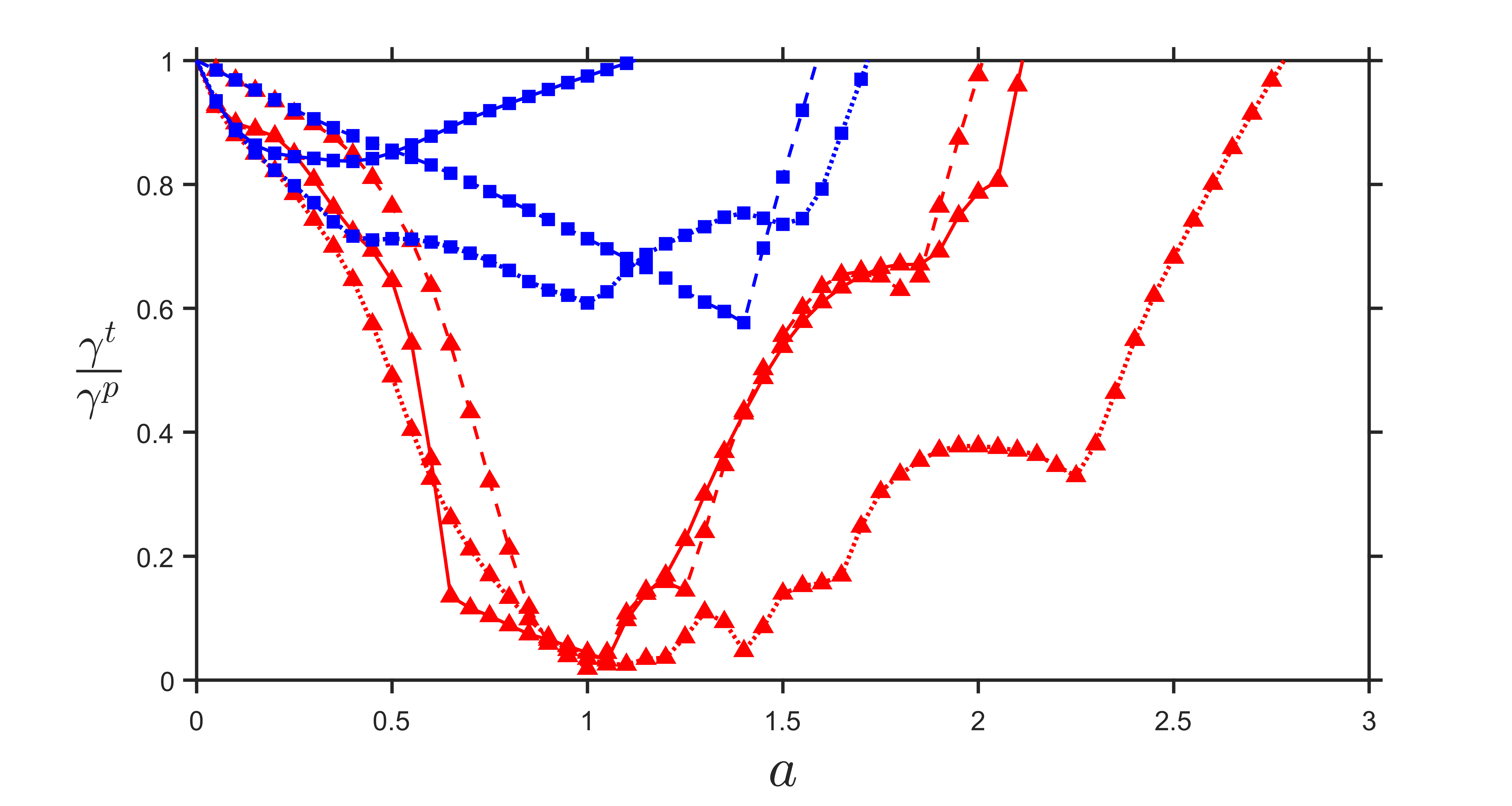}
\caption{Tertiary growth rate $\gamma^t$, normalised to the primary growth rate $\gamma^p$, of different zonal flow profiles obtained from nonlinear simulations as their amplitude is multiplied by a factor $a$. Quiescent zonal profiles (red triangles) are seen to typically be of very nearly that amplitude which is most stabilising, i.e. when $a=1$. The same is not true for profiles from turbulent periods (blue squares), which, beyond generally being less stabilising, may be made more stabilising if their amplitude is altered.}
\label{FigureTrendingToStabilisation}
\end{figure}

Having observed that the quasistationary zonal flows are very nearly tertiary stable, the question becomes how robust this stability is. For the fluid model under consideration in the previous work of \citet{Hallenbert2021} this concept was key. There, frequent turbulent bursts sustained drift waves at sufficient amplitude to affect stable zonal profiles while decaying, markedly altering them. Therefore, to prevent subsequent turbulent bursts, the zonal profile had to not only be individually tertiary stable, but also belong to a family of similar profiles that also were stable. The need for such stringent stability is presently much lesser. As observed in figure \ref{FigureZonalEvolution}, the Z-pinch exhibits very little drift wave activity over extended periods in the Dimits regime, during which the zonal flow remains very nearly unchanged. As such, individual profile stability is therefore re-emphasised.

In spite of what was just discussed, investigating the question of ``stability robustness" is still of interest. In figure \ref{FigureTrendingToStabilisation}, therefore, the maximal $\gamma^t$ is plotted for various zonal flow profiles, versus the rescaling factor $a$ used to (artificially) adjust the zonal flow amplitude. Clearly, $\gamma^t$ increases relatively swiftly away from marginal stability as $a$ is moved away from $1$ for quasistationary profiles. On the other hand, the turbulent profiles can be made somewhat more stable for some values $a\neq1$, but never to the same extent as the quasistationary ones. We interpret this as meaning that quasistationary flows, in being established when turbulent bursts end, are delicately pushed towards configurations of greater tertiary stability, all the way to marginal stability if possible. We also note again that increasing the zonal shear by increasing $a$ does not necessarily imply that $\gamma^t$ decreases, a fact already noted by \citet{Kobayashi2012}, who however misattributed this fact to KH-type zonal flow breakup instead of their reduced efficacy in stabilising drift waves, something we will expand upon in \S\;\ref{SectionAdaptation}.

\subsection{Tertiary instability of a single zonal mode}\label{SectionSingleMode}

To gain some basic understanding of the tertiary instability, we now turn to the case of a single zonal mode i.e. a sinusoidal zonal profile, and in particular the 4M-reduced system obtained when $q_\mathrm{G}=q$ in \eqref{CapitalQ}, which is the minimal system that can encapsulate zonal shear stabilisation. As is apparent from the presence of $\overline{g}_s$ in $\mathsfbi{C}_2$ the zonal flow $\partial_x\overline{\varphi}$ alone is insufficient to determine tertiary instability; the specific form of the zonal electron/ion distribution giving rise any particular zonal flow also matters by modifying the sideband coupling. Naturally, one reason for this is because $\overline{g}_s$ in effect act as a modification to the background gradients. However, its influence is not restricted to this effect since the full kinetic zonal distribution $g_{s\boldsymbol{q}}$ has to be accounted for, not just its density and temperature moments. This fact, necessarily not included in previous studies of fluid systems like \citet{Zhu2020}, \citet{Ivanov2020}, or \citet{Hallenbert2021}, has perhaps not been adequately appreciated. To investigate this effect in a controlled way, we will in the next sections vary the zonal distribution $g_{s\boldsymbol{q}}$, holding $\overline{\varphi}_{\boldsymbol{q}}$ fixed by employing a multiplicative factor.

Before delving into this topic, we will first discuss a way in which $\overline{g}_s$ does not affect the tertiary instability. In \hyperref[AppendixInvariance]{Appendix \ref*{AppendixInvariance}} we show that the tertiary growth rate of a single zonal mode is unchanged under the transformation 
\begin{equation}
    \overline{g}_{s\boldsymbol{q}} \rightarrow  \frac{\overline{g}_{s\boldsymbol{q}}^*\overline{\varphi}_{\boldsymbol{q}}^2}{|\overline{\varphi}_{\boldsymbol{q}}|^2}.
\end{equation} 
Since this substitution also leaves $\overline{\varphi}$ unchanged, the tertiary instability can only possess a $\overline{g}_{s\boldsymbol{q}}$-dependence of the (species-summed) form $|g_{s\boldsymbol{q}}|$ and $g_{s\boldsymbol{q}}+g_{s\boldsymbol{q}}^*\overline{\varphi}_{\boldsymbol{q}}^2/|\overline{\varphi}_{\boldsymbol{q}}|^2$, i.e. $\Re{(g_{s\boldsymbol{q}}\overline{\varphi}_{\boldsymbol{q}}}^*)$. As a consequence, it does not matter which species distribution trails the potential and which leads it. This result can be interpreted in light of the tertiary localisation highlighted in \S\;\ref{SectionLocalisation}. The part of the particle distribution $g_{s\mathrm{out}}$ that is out of phase with $\overline{\varphi}$ satisfies $\partial_xg_{s\mathrm{out}}=0$ at the points $\partial^2_x\overline{\varphi}=0$ that are tertiary unstable, and so are not able to affect the tertiary instability by modifying the background gradients there.

\begin{figure}
\centering
\includegraphics[width=\linewidth]{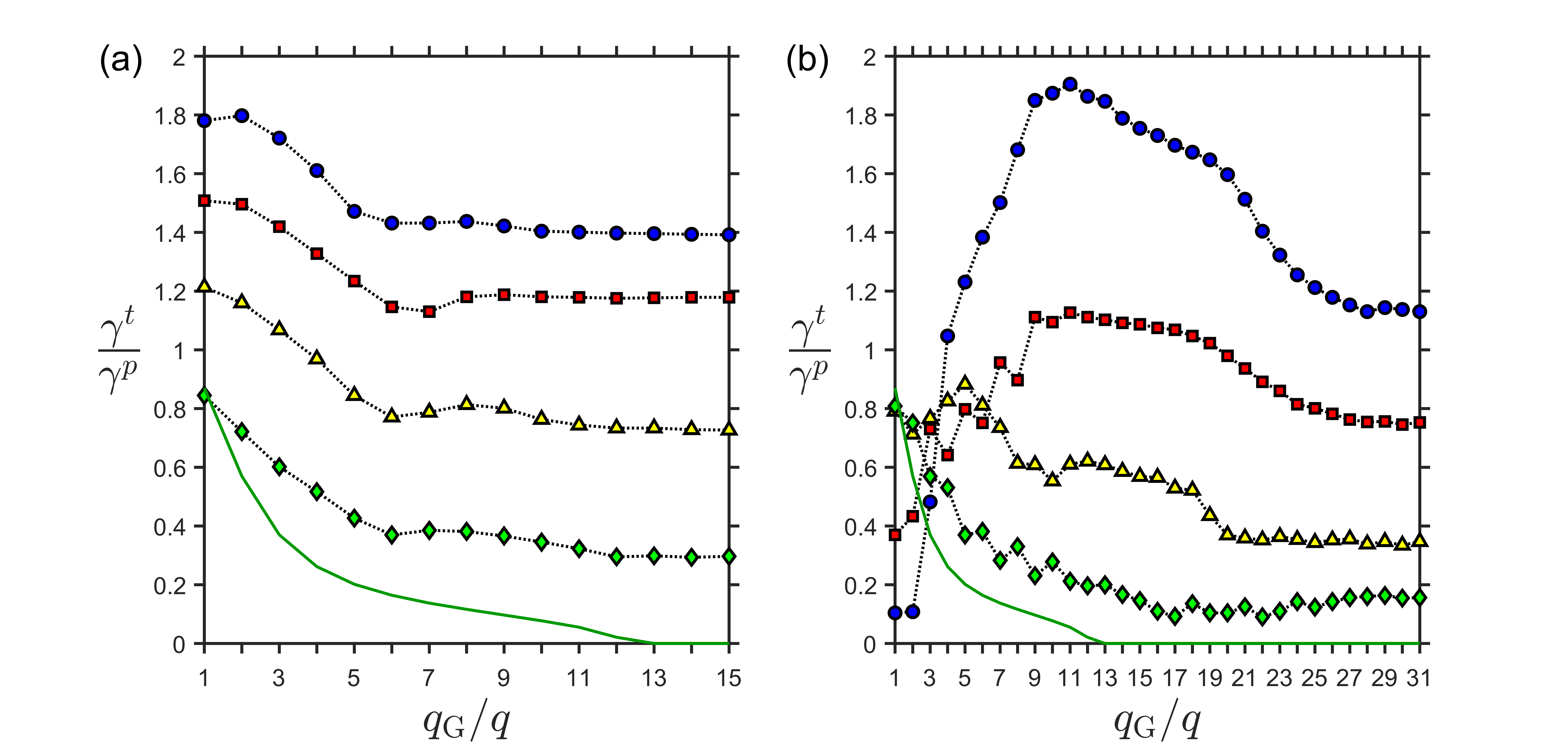}
\caption{Tertiary growth rate $\gamma^t$ at $\eta=0.25$ and $R/L_n=1.8$ for $p=0.6$, normalised to the primary growth rate $\gamma^p$, of a single zonal mode with wavenumber $q=0.4$ for different example configurations of the form \eqref{Equation4MDistributions} for (a) $\overline{\varphi}_{\boldsymbol{q}}$=3.5, $\alpha_1+\alpha_2 i=(0.5, 2+i, 0, -1)$, and (b) $\overline{\varphi}_{\boldsymbol{q}}$=35, $\alpha_1+\alpha_2 i=(1+i, 0.4+i, 0.1+0.25i, -1+i)$. Also plotted in green is the primary growth rate $\gamma^p_{\boldsymbol{p}+\boldsymbol{q}_\mathrm{G}}$ of the sideband mode with $k_x=q_\mathrm{G}$. The configurations are chosen to give a wide spread of converged $\gamma^t$, which occurs around $q_\mathrm{G}/q\sim10$ and $q_\mathrm{G}/q\sim30$ for $\overline{\varphi}_{\boldsymbol{q}}=3.5$ and $\overline{\varphi}_{\boldsymbol{q}}=35$ respectively, beyond which point further coupling to smaller scale sidebands becomes negligible. As is shown, the growth rate decreases as smaller scale sidebands are engaged, but not below $\gamma^p_{\boldsymbol{p}+\boldsymbol{q}_\mathrm{G}}$. For greater zonal amplitudes, however, both patterns are disrupted for small $q_\mathrm{G}$.}
\label{FigureModesConvergence}
\end{figure}

The preceding discussion did not depend upon the Galerkin truncation. However, before we further investigate the role of $g_{s\boldsymbol{q}}$ in \S\;\ref{SectionElectronIonPhase} for the tertiary instability, it is prudent to investigate its effect. Indeed, the single zonal mode presently under consideration is suitable for this purpose. Thus, how $\gamma^t$, precisely obtained with GENE as outlined in \S\;\ref{SectionTertiarySimulations}, changes as $q_\mathrm{G}$ is varied for some different single-mode configurations with $q=0.4$, $p=0.6$, and $\overline{\varphi}_{\boldsymbol{q}}=3.5$ (a typical nonlinear value) or $\overline{\varphi}_{\boldsymbol{q}}=35$, is compared with the primary sideband growth rate $\gamma^p_{\boldsymbol{p}+\boldsymbol{q}_\mathrm{G}}$ in figure \ref{FigureModesConvergence}. For small to ``normal" zonal amplitudes $\gamma^t>\gamma^p_{\boldsymbol{p}+\boldsymbol{q}_\mathrm{G}}$ seems to uniformly hold in accordance with the view that the tertiary instability mixes primary modes \citep{Hallenbert2021}. When significantly increased,
$\gamma^t>\gamma^p_{\boldsymbol{p}+\boldsymbol{q}_\mathrm{G}}$ may however hold for small $\boldsymbol{q}_\mathrm{G}$, a feature presumably explained by immediate subdominant stable mode coupling \citep[see e.g][]{Terry2006,Hatch2011}, the matching of which becomes less resonant when more modes are included, i.e. when $\boldsymbol{q}_\mathrm{G}$ is increased. 

Continuing, we note that the point at which $\gamma^t$ converges effectively constitutes a measure of how many sidebands are coupled together into a combined tertiary mode, which intuitively increases with increasing $\overline{\varphi}_{\boldsymbol{q}}$, since in the opposite weak coupling limit $\overline{\varphi}_{\boldsymbol{q}}\rightarrow0$ different wavenumbers decouple. For the typical zonal amplitude of figure \ref{FigureModesConvergence} (a) this point occurs at $k_x\approx12q=4.8<6.4$, which as stated in \S\;\ref{SectionNonlinearSimulation} is the smallest scale included in our simulations, offering some further evidence that our numerical implementation was sufficient to capture the dynamics of the Dimits regime.

\subsection{Role of electron/ion response phase}\label{SectionElectronIonPhase}

Leaving aside the complicated question of how the inclusion of multiple zonal modes with their accompanying cross-phases affect the tertiary instability, we proceed to investigate how important the different zonal distribution effects uncovered in \S\;\ref{SectionSingleMode} are, the first of which is the relative phase of the ion/electron response. To investigate this feature, both distributions are taken to possess a simple representative Maxwellian $f_{0s}$-dependence, i.e.
\begin{equation}\label{Equation4MDistributions}
     g_{e\boldsymbol{q}} = M(\alpha_1+\alpha_2 i)\left(\frac{m_e}{m_i\pi\alpha_3}\right)^{3/2}e^{-m_ev^2/\alpha_3m_i} \;\;\; \mathrm{and} \;\;\; g_{i\boldsymbol{q}} = \frac{M}{(\pi\alpha_4)^{3/2}}e^{-v^2/\alpha_4}
\end{equation}
where $M$ is a multiplicative factor that ensures $\overline{\varphi}_{\boldsymbol{q}}$ attains the desired value $\overline{\varphi}_{\boldsymbol{q}\mathrm{des}}$, and which can be calculated using quasineutrality \eqref{quasineutrality} as
\begin{equation}
    M = \frac{\overline{\varphi}_{\boldsymbol{q}\mathrm{des}}(2-\Gamma_{0i\boldsymbol{q}}-\Gamma_{0e\boldsymbol{q}})}{\int d^3 vJ_{0i\boldsymbol{q}}g_{i\boldsymbol{q}}/M - \int d^3 vJ_{0e\boldsymbol{q}}g_{e\boldsymbol{q}}/M}.
\end{equation}
Here 
\begin{equation}
    \frac{ n_{ge\boldsymbol{q}} }{ n_{ge\boldsymbol{q}}} = \frac{\int d^3v g_{e\boldsymbol{q}} }{ \int d^3v g_{i\boldsymbol{q}}} = \alpha_1+\alpha_2i 
\end{equation} 
is the cross-species phase whose influence we want to investigate. On the other hand, $\alpha_3=\alpha_4=0.2$ are ``effective temperatures", the choice of which will be explained in \S\;\ref{SectionPrediction}. 

\begin{figure}
\centering
\includegraphics[width=0.95\linewidth]{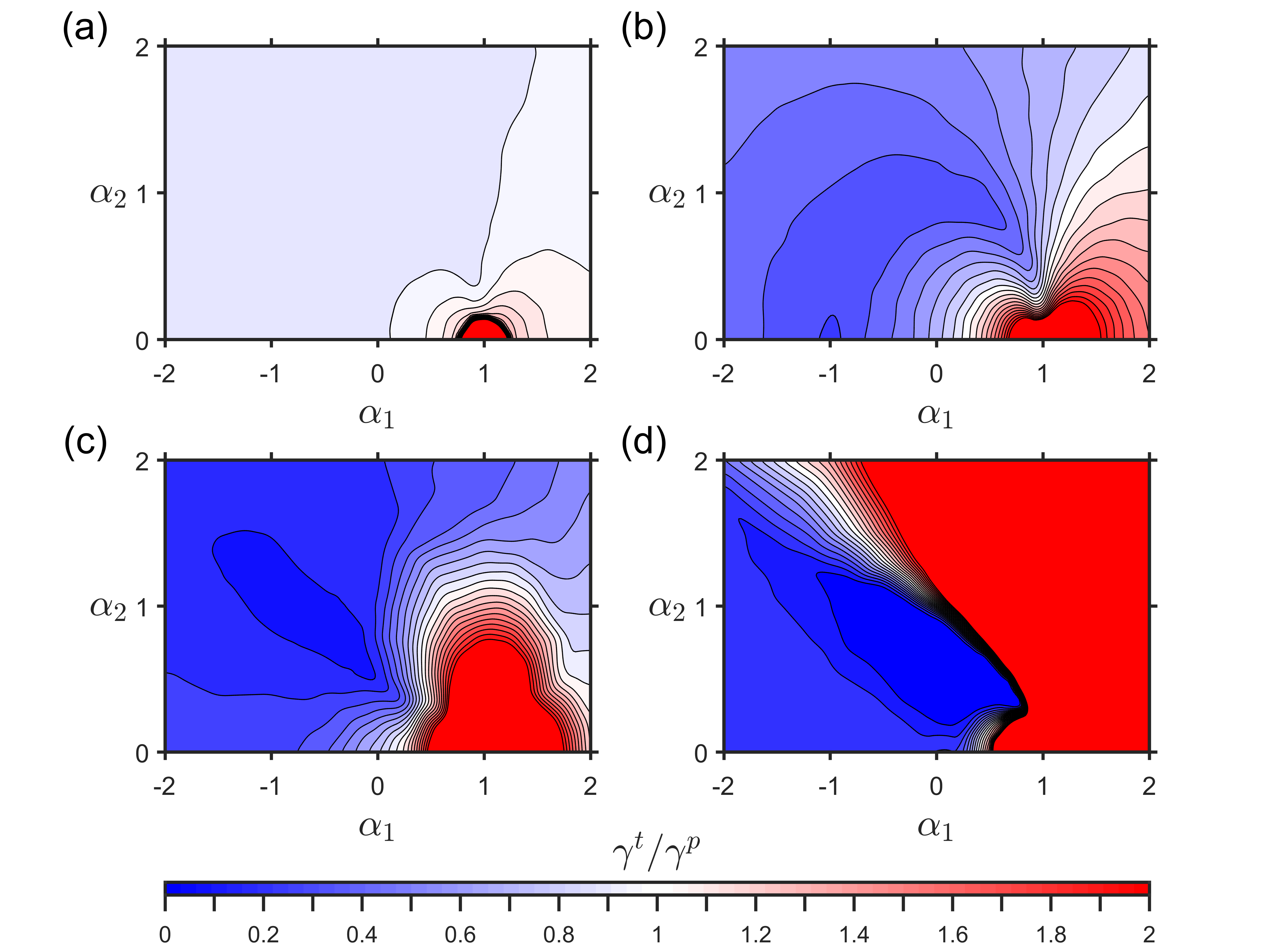}
\caption{Tertiary instability growth $\gamma^t$ for $\eta=0.25$, $R/L_n=1.8$, and $p=0.6$ in the presence of a sinusoidal zonal profile of $q=0.4$ and amplitude (a) $\overline{\varphi}_{\boldsymbol{q}}=0.35$, (b) $\overline{\varphi}_{\boldsymbol{q}}=3.5$, (c) $\overline{\varphi}_{\boldsymbol{q}}=35$, and (d) $\overline{\varphi}_{\boldsymbol{q}}=350$, as a function of the relative phase $\alpha_1+\alpha_2i=\int d^3v g_{e\boldsymbol{q}} / \int d^3v g_{i\boldsymbol{q}}$ of the zonal electron/ion responses of \eqref{Equation4MDistributions}. Though the stabilisation changes significantly as the zonal amplitude is varied, two noteworthy features persist: maximum instability occurs for $(\alpha_1,\alpha_2)=(1,0)$; the case $(\alpha_1,\alpha_2)=(-1,0)$, though not necessarily most stabilising, is always very stabilising.}
\label{FigureElectronIonResponse}
\end{figure}
 
Figure \ref{FigureElectronIonResponse} shows how $\gamma^t$ of the poloidal band $p=0.6$ at the Dimits threshold $\eta=0.25$ and $R/L_n=1.8$ changes as $\alpha_3/\alpha_4$ are varied, Four different cases, corresponding to $\varphi_{\boldsymbol{q}}=0.35, 3.5, 35$, and $350$, are shown, each exhibiting a radically different appearance (note the usage of the result of \hyperref[AppendixInvariance]{Appendix \ref*{AppendixInvariance}} for a reduction to the half-plane). Two features, however, persist. The first is an extreme instability $\gamma^t\gg\gamma^p$, centred at the point $(\alpha_1,\alpha_2)=(1,0)$ where the ion/electron responses are completely in phase and which spreads with increasing $\overline{\varphi}_{\boldsymbol{q}}$. The second feature is the stabilisation $\gamma^t<\gamma^p$ of out-of-phase responses with $\alpha_1<0$ so long as $\alpha_2$ is not too large.

\begin{figure}
\centering
\includegraphics[width=1.05\linewidth]{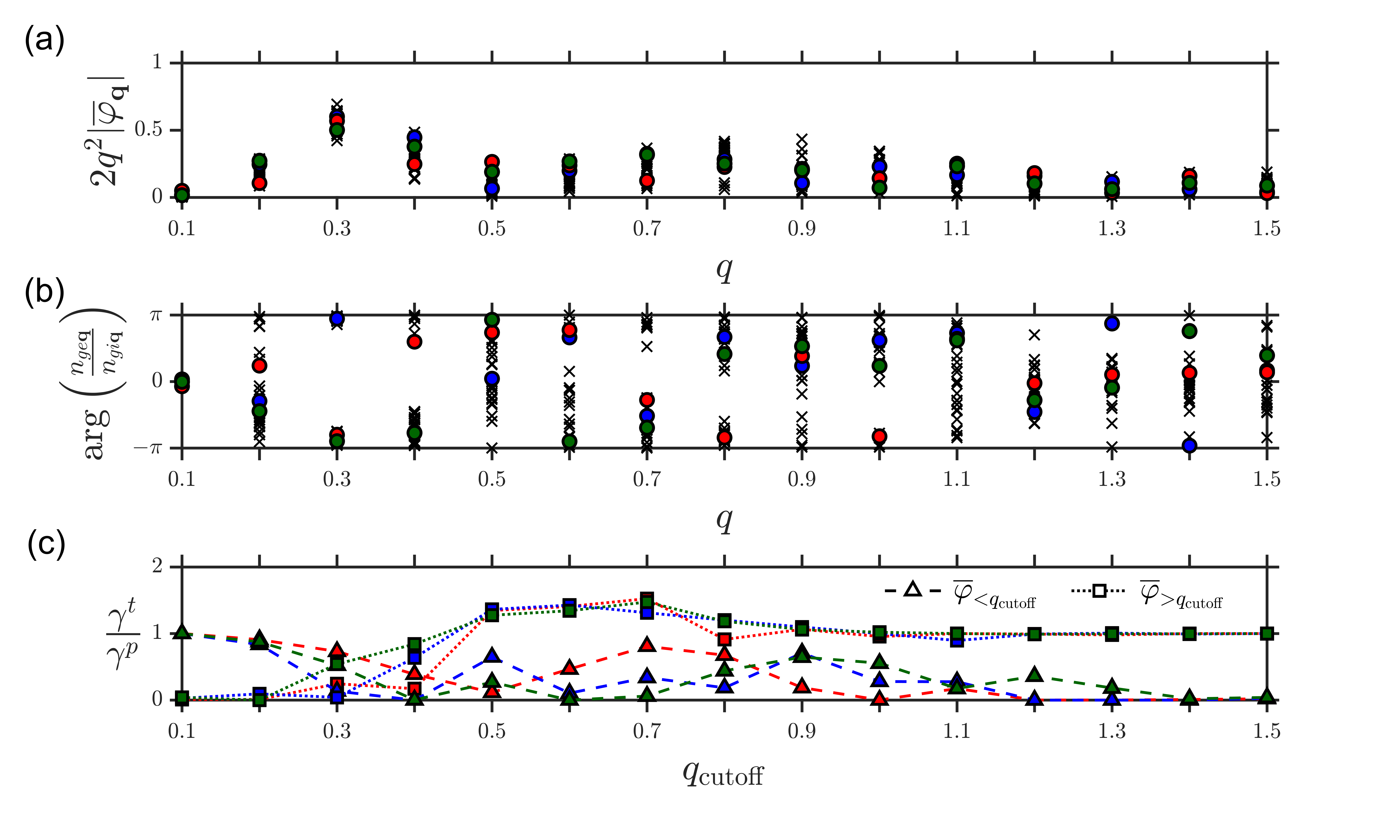}
\caption{(a) The individual mode contribution $2q^2|\overline{\varphi}_{\boldsymbol{q}}|$ to the squared box-averaged zonal shear $\langle|\partial_x^2\overline{\varphi}|^2\rangle$ of different zonal quiescent zonal profiles obtained from nonlinear simulations with $(\eta,R/L_n)$ closely clustered around $(0.25,1.7)$, three of which (red, blue, green) are highlighted. (b) the argument of the corresponding relative phase $n_{ge\boldsymbol{q}}/n_{gi\boldsymbol{q}}$. (c) the corresponding tertiary growth rates for $p=0.6$ of the modified profiles $\overline{\varphi}_<$ and $\overline{\varphi}_>$ that are obtained via the removal of modes not satisfying $q<q_\mathrm{cutoff}$ (triangles) and $q>q_\mathrm{cutoff}$ (squares) respectively. It is apparent that the modes with $q=0.3$ and $q=0.4$ typically develops a phase shift of $\sim\pi$ and thus play an inordinately important role (compared to their shearing) for the profile's total stabilising ability, since $\gamma^t$ greatly increases around $q_\mathrm{cutoff}\sim0.3-0.4$ when large scale modes are excluded. When instead small scale modes are excluded, $\gamma^t$ instead decreases there, but continues to vary significantly and grow large again beyond this point, showing that smaller scale modes still can play an important role.}
\label{FigurePhaseShift}
\end{figure}

Based on the above result and the findings of \S\;\ref{SectionTertiaryInstability}, we conclude that quiescent zonal profiles should exhibit this ion/electron phase shift, which the zonal temperature profiles of figure \ref{FigureTertiaryLocalisation} already hinted at. To firmly establish that this is the case, multiple simulations scattered closely around the point $\eta=0.25$ and $R/L_n=1.7$ were conducted to extract different long-term stable zonal distributions. The result of these can be seen in figure \ref{FigurePhaseShift}, where the distribution of (a) the contribution $2q^2|\overline{\varphi}_{\boldsymbol{q}}|$ to the zonal shear and (b) $\mathrm{arg}(n_{ge\boldsymbol{q}}/n_{ge\boldsymbol{q}})$ for the first $15$ zonal modes is plotted. It is striking that the $q=0.3$ (whose shear contribution typically is the largest) and $q=0.4$-modes indeed develops the expected phase shift $\mathrm{arg}(n_{ge\boldsymbol{q}}/n_{ge\boldsymbol{q}})\approx\pi$, but confusingly other modes are seen to exhibit similar ``avoidance zones" centred elsewhere or appear nearly uniformly distributed.

The above picture is clarified when we selectively filter out Fourier components to define high- and low-pass-filtered zonal fields $\overline{\varphi}_{>q_\mathrm{cutoff}}$ and $\overline{\varphi}_{<q_\mathrm{cutoff}}$. The tertiary modes are then simulated with GENE and the resulting growth rate $\gamma^t$ of three such profiles shown in \ref{FigurePhaseShift} (c). $\gamma^t$ of the low-pass-filtered fields is seen to rapidly increase from $\sim0$ to greater than $\gamma^p$ where it then remains as $q_\mathrm{cutoff}$ goes from $0.3$ to $0.5$, i.e. when the aforementioned $q=0.3$ and $q=0.4$-modes are excluded. For the high-pass-filtered fields, a similar trend is observed when $q_\mathrm{cutoff}$ traverses in the opposite direction, though here $\gamma^t$ also exhibits fluctuating large values also in the range $0.5\gamma^p-1.2\gamma^p$. Combined, it therefore seems that the phase-shifted modes contributes especially strongly to the total stabilisation compared with other modes, as one would suspect based solely on their shear. These modes can be thought of as acting as a foundation to which modes at smaller scale modes can contribute, constructively or destructively, but without which smaller scale modes fail to even approach stability; in short, they are the most important component.

\subsection{Velocity space dependence}\label{SectionVelocitySpace}

Another dependence of the tertiary we noted in \S\;\ref{SectionSingleMode} was that the dependence upon velocity space moments and so we wish to explore how $\gamma^t$ depends upon 
\begin{equation}\label{VMoments}
    \int d^3vv^{2n}g_{s\boldsymbol{q}}, \;\;\; n\in\{1,2,3,...\},
\end{equation}
for some representative $g_{s\boldsymbol{q}}$-functions. Naturally, we know from \S\;\ref{SectionSingleMode} that the relevant velocity moments for the tertiary theory have the form
\begin{equation}
    \int d^3v\frac{v^{2n}g_{s\boldsymbol{q}}}{L_s}, \;\;\; n\in\{1,2,3,...\},
\end{equation}
where the resonant denominator $L_s$ originates from $\mathsfbi{A}_s^{-1}$. Unfortunately, $L_s$ will generally depend upon $\lambda^t$ and $v^2$, which necessitates the solution of an intractable integral equation for a rigorous determination of the velocity moments controlling the tertiary mode. We therefore focus on the general moments \eqref{VMoments} to obtain a rough assessment of the importance of velocity space dependence.

Proceeding by lifting the most stable distributions of the form \eqref{Equation4MDistributions} in figure \ref{FigureElectronIonResponse} (b) and (c), we then add an additional part $g^T_{s\boldsymbol{q}}=p_1(v_\bot^2)\exp\left(-v^2\right)$, where $p_1$ is the third degree polynomial satisfying 
\begin{equation}
    \int d^3vv_\bot^2p_1(v_\bot^2)e^{-v^2}=1,\;\;\; \mathrm{and} \;\;\; \int d^3vp_1(v_\bot^2)e^{-v^2}=\int d^3vv_\bot^4p_1(v_\bot^2)e^{-v^2}=0,
\end{equation}
or 
$g^\chi_{s\boldsymbol{q}}=p_2(v_\bot^2)\exp\left(-v^2\right)$, where $p_2$ is the third degree polynomial satisfying 
\begin{equation}
    \int d^3vv_\bot^4p_2(v_\bot^2)e^{-v^2}=1\;\;\; \mathrm{and} \;\;\; \int d^3vp_2(v_\bot^2)e^{-v^2}=\int d^3vv_\bot^2p_2(v_\bot^2)e^{-v^2}=0.
\end{equation}
These additions should be thought of as a pure temperature perturbation and a purely higher moment perturbation respectively, whose stabilisation effect we are interested in.

\begin{figure}
\centering
\includegraphics[width=\linewidth]{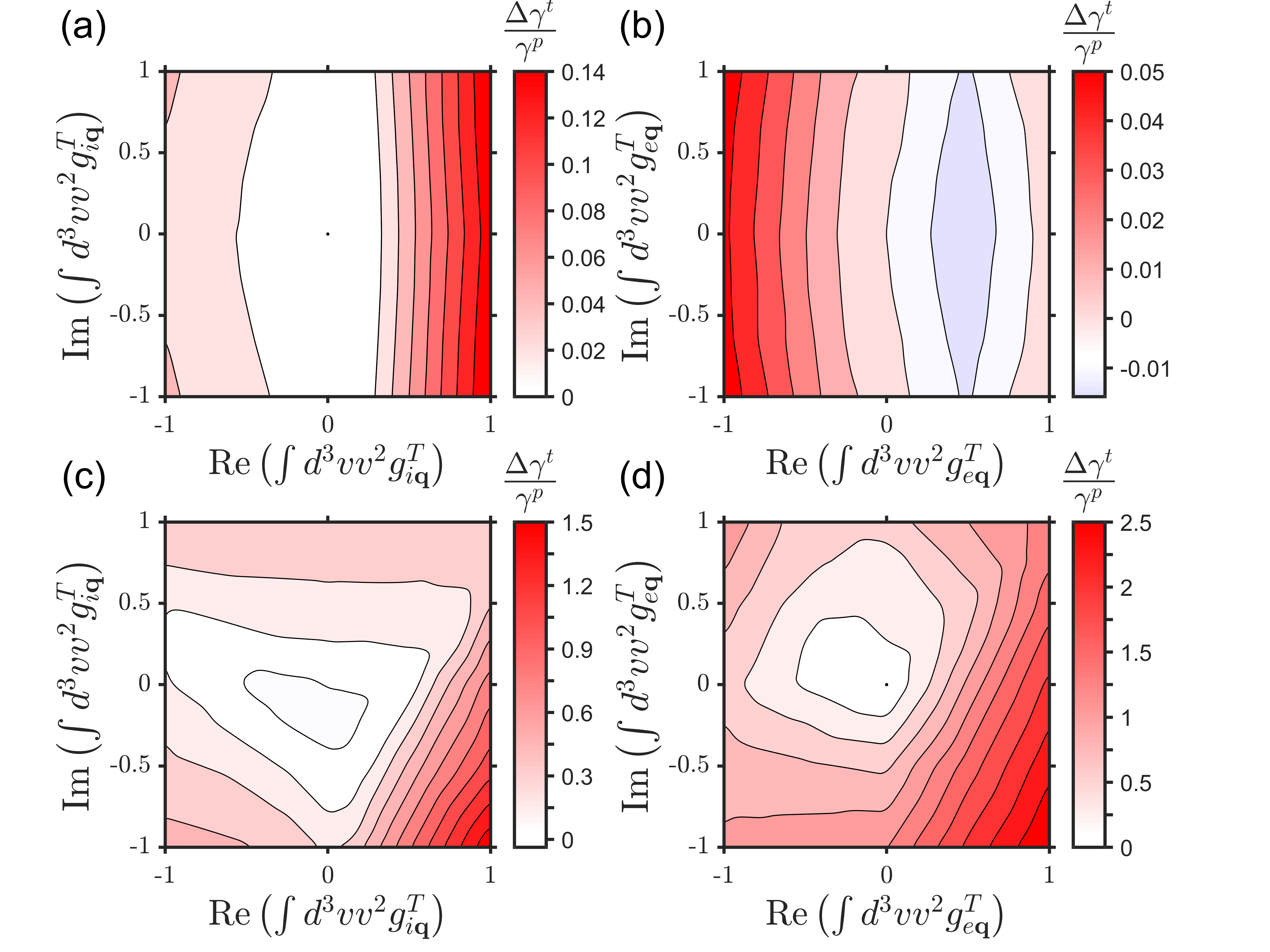}
\caption{Tertiary growth rate modification $\Delta\gamma^t$ when an additional part $g^T_{s\boldsymbol{q}}=p_1(v_\bot^2)\exp\left(-v^2\right)$, with $p_1(v_\bot^2)$ being the second degree $v_\bot^2$-polynomial such that $g^T_{s\boldsymbol{q}}$ has vanishing zeroth and fourth velocity moments but non-vanishing second velocity moment, is added to the initial, pre-renormalisation zonal response $g^n_{s\boldsymbol{q}}$ employed in figure \ref{FigureElectronIonResponse}. Specifically, for (a) and (b) $(\alpha_1,\alpha_2,\overline{\varphi}_{\boldsymbol{q}})=(-1,0,3.5)$, corresponding to the most stable point of figure \ref{FigureElectronIonResponse} (b) where $\overline{\varphi}_{\boldsymbol{q}}=3.5$, while for (c) and (d) $(\alpha_1,\alpha_2,\overline{\varphi}_{\boldsymbol{q}})=(0,0.5,35)$, corresponding to the most stable point of figure \ref{FigureElectronIonResponse} (c) where $\overline{\varphi}_{\boldsymbol{q}}=35$. The inclusion of the ``pure temperature perturbation" $g^T_{s\boldsymbol{q}}$ is seen to predominantly destabilise the tertiary instability, and when it is stabilising its effect does not exceed $\sim1-2\%$ of the primary growth rate $\gamma^p$.}
\label{FigureTemperatureMomentResponse}
\end{figure}

\begin{figure}
\centering
\includegraphics[width=\linewidth]{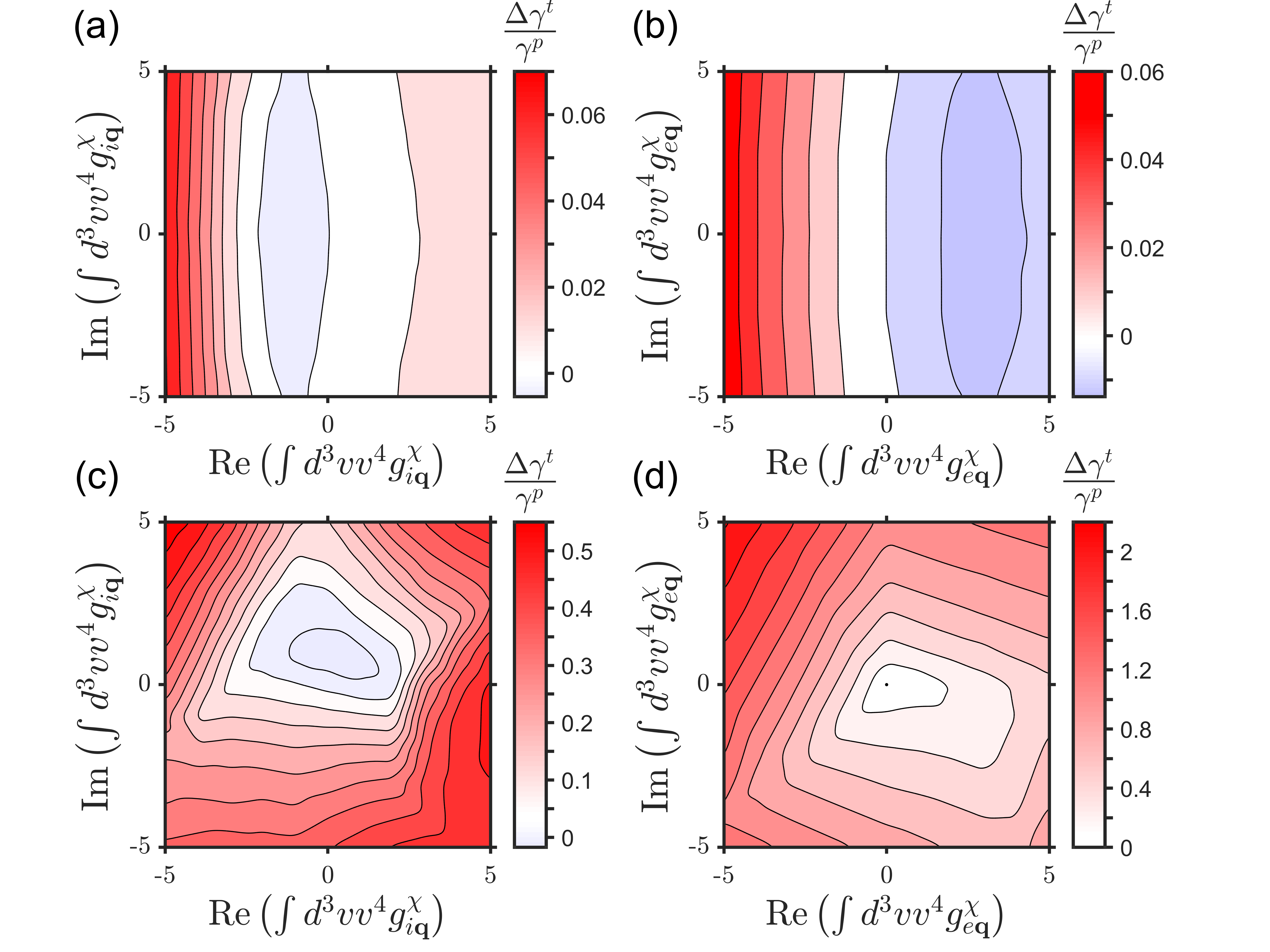}
\caption{The equivalent result to figure \ref{FigureTemperatureMomentResponse}, but where instead $g^\chi_{s\boldsymbol{q}}=p_2(v_\bot^2)\exp\left(-v^2\right)$ with $p_2(v_\bot^2)$ being the second degree $v_\bot^2$-polynomial such that $g^T_{s\boldsymbol{q}}$ has vanishing zeroth and second velocity moments but non-vanishing fourth velocity moment. Once again, the inclusion of a higher moment perturbation is seen to predominantly destabilise the tertiary instability, with marginal stabilisation not exceeding $\sim2\%$. }
\label{FigureHigherMomentResponse}
\end{figure}

Now, the resulting degree $\Delta\gamma^t$ that the tertiary growth rate $\gamma^t$ is modified by the presence of $g_{s\boldsymbol{q}}^T$ and $g_{s\boldsymbol{q}}^\chi$ can be seen in Figures \ref{FigureTemperatureMomentResponse} and \ref{FigureHigherMomentResponse}. It is very clear that in all cases, even when tuned correctly, the inclusion of either term could only marginally reduce $\gamma^t$ by some $0.01\gamma^p$. On the other hand, for other values that are not tuned the net effect of said inclusion rapidly becomes strongly destabilising. Of course we cannot guarantee that it is impossible to employ some trial function that will significantly stabilise $\gamma^t$. Nevertheless a range of different exploratory investigations with different trial functions hints that the velocity space stabilisation cannot move $\gamma^t$ much at all below its minimum value among the configurations of figure \ref{FigureElectronIonResponse}.

\section{A reduced mode tertiary instability Dimits shift prediction}\label{SectionPrediction}

At last we proceed to the motivation behind the present work: attempting to expand and apply the reduced mode Dimits shift estimate outlined in \citet{Hallenbert2021}, which proved successful in the gyrokinetic strongly driven fluid limit system. Expressed in the present notation, but introducing a $4M$-subscript to explicitly show that a ``reduced" 4M truncated tertiary instability is considered, the predicted Dimits threshold was obtained from the critical solution of the equation system involving the four system parameters $p$, $q$, $\overline{\varphi}_{\boldsymbol{q}}$, and $R/L_n$ that is given by
\begin{equation}\label{cond1}
    \frac{\partial \gamma^{p}_{\boldsymbol{p}}}{\partial p}=0,
\end{equation}
\begin{equation}\label{cond2}
    \frac{\partial\gamma^{t}_{4M}}{\partial \overline{\varphi}_{\boldsymbol{q}}}=0,
\end{equation}
\begin{equation}\label{cond3}
    \frac{\partial \gamma^{t}_{4M}}{\partial q}=0,
\end{equation}
\begin{equation}\label{cond4}
    \gamma^{t}_{4M}\left(\frac{R}{L_n}\right) = 0.
\end{equation}
Here the last equation is the ``most important", as its solution specifies a $L_n$-value that is taken to correspond to the Dimits threshold.

The reason why the 4M-reduction was employed essentially reduced to the fact that it greatly increases computational efficiency, a very valuable feature in increasingly complex systems. Additionally, a 4M prediction was there found to be lower than a non-truncated prediction. This was advantageous since it was observed that zonal profiles characterising that Dimits range typically were ``robustly stable" to small-amplitude drift waves rather than merely tertiary stable. The lower 4M prediction could thus be thought to approximate this effect, but may not prove generally suitable. 

We will now go through Equations \eqref{cond1}-\eqref{cond4} to make their meaning clear. Starting from the bottom, Equation \eqref{cond4} succinctly expresses that the Dimits threshold corresponds to a tertiary stability threshold, specifically one which is captured by the 4M-reduced system. Equation \eqref{cond1} next conveys the expectation that the tertiary mode of the poloidal band containing the most unstable primary mode also is the most unstable, and thus destabilises first. The single zonal mode in the system, and whose instability threshold is taken to approximate a typical full zonal profile, is then set through Equations \eqref{cond2} and \eqref{cond3}, where the former fixes its amplitude and the latter its wavelength, both so that the tertiary mode is maximally stable. 

Equations \eqref{cond1}-\eqref{cond4} taken together can thus be summarised as the conditions for marginal stability, \eqref{cond4}, of the most unstable tertiary mode, \eqref{cond1}, assuming a maximally stabilising sinusoidal zonal profile, \eqref{cond2} and \eqref{cond3}.  This, as explained in \citet{Hallenbert2021}, attempts to capture how the random nature of turbulence makes it plausible that the system will continue to explore the landscape of zonal flows until it finds a zonal profile that, if it exists, is able to stabilise the system. Particularly, since energy will be injected at a faster rate for unstable profiles, the system will then be more free, and thus prone, to evolve into a different configuration. 

Rather than accepting the above description wholesale, it is prudent to again review its applicability and validity. To this end we recall that, as indicated by Figures \ref{FigureZonalEvolution}, \ref{FigureTertiaryKy}, and \ref{FigureTrendingToStabilisation}, $\gamma^t$ is marginal for the quasistationary flows of the Dimits regime and increases substantially above it. Thus it seems the transition is indeed characterised by some condition like \eqref{cond4}. Similarly, figure \ref{FigureTertiaryKy} makes it patently clear that \ref{cond1} is a suitable and robust condition. That leaves just the 4M-reduction to a single zonal mode of \eqref{cond2} and \eqref{cond3}, but once again, as a result of the observation of figure \ref{FigurePhaseShift} that a few modes gives rise to most stabilisation and the evolution towards stable flows entailed by figure \ref{FigureTrendingToStabilisation}, it is hinted that a reduced mode description should prove apt.

\subsection{Adapting the prediction}\label{SectionAdaptation}

Now, the astute reader will clearly have identified the problem with Condition \eqref{cond2} as written: formulated for a fluid model it does not prescribe how one should choose $g_{s\boldsymbol{q}}$. Of course, this is in principle easily fixed by extending the optimisation to find the maximally stabilising distributions, but straightforwardly doing so entirely eliminates the simplicity and computational efficiency of the prediction with the vast possibility space. We therefore reduce our attention to some trial functions
\begin{equation}\label{EquationTrialFunctions}
    g_{s\boldsymbol{q}}=h(\alpha_{s1},\alpha_{s2},...,\alpha_{sn_s},w_\bot^2,w_\parallel^2),
\end{equation}
with $n_e+n_i$ variable parameters $\alpha_{si}$ with $i\in1,2,...,n_s$, the determination of whom is given by
\begin{equation}\label{cond5}
    \frac{\partial\gamma^t}{\partial\alpha_{si}}=0,
\end{equation}
which are added to Conditions \eqref{cond1}-\eqref{cond4} to complete our system. Here we must obviously attempt to find as minimal a set as possible that we can still trust to lead us reasonably close to the optimal configuration. 

Now with a drive term proportional to $f_{0s}(T)$, it is natural to assume a similar Maxwellian velocity-dependence, but an initial, exploratory investigation revealed that $f_{0s}(bT)$ with $b\approx0.2$ generally yielded much lower tertiary growth rates. This proved consistent enough that, though included as an optimisation parameter, it could safely be left at this value with minimal impact for typical values $q\lesssim p$ and $\overline{\varphi}_{\boldsymbol{q}}\sim 1-10$. Beyond this, the analysis of \S\;\ref{SectionElectronIonPhase}, and figure \ref{FigureElectronIonResponse} in particular, makes it clear that the cross-species phase $n_{ge\boldsymbol{q}} / n_{gi\boldsymbol{q}}$ is a vital parameter to include. On the other hand, \S\;\ref{SectionVelocitySpace} strongly hints that, when searching for stable distributions, the inclusion of finer velocity space structures in our functional dependence can safely be neglected at an accuracy cost of perhaps only some $0.02\gamma^p$; their propensity to cause instability makes them, perhaps counter-intuitively, unimportant for the Dimits shift, since in the Dimits regime, where zonal flows evolve towards stability, they are unfavoured to emerge. This is a massively favourable result for the present prediction, and means that we need not search for further variable parameters. Thus we will employ \eqref{Equation4MDistributions} in \eqref{EquationTrialFunctions} with $\alpha_3=0.2$, thereby restricting \eqref{cond5} to just $\alpha_1$ and $\alpha_2$, which should prove sufficient.

\begin{figure}
\centering
\includegraphics[width=0.8\linewidth]{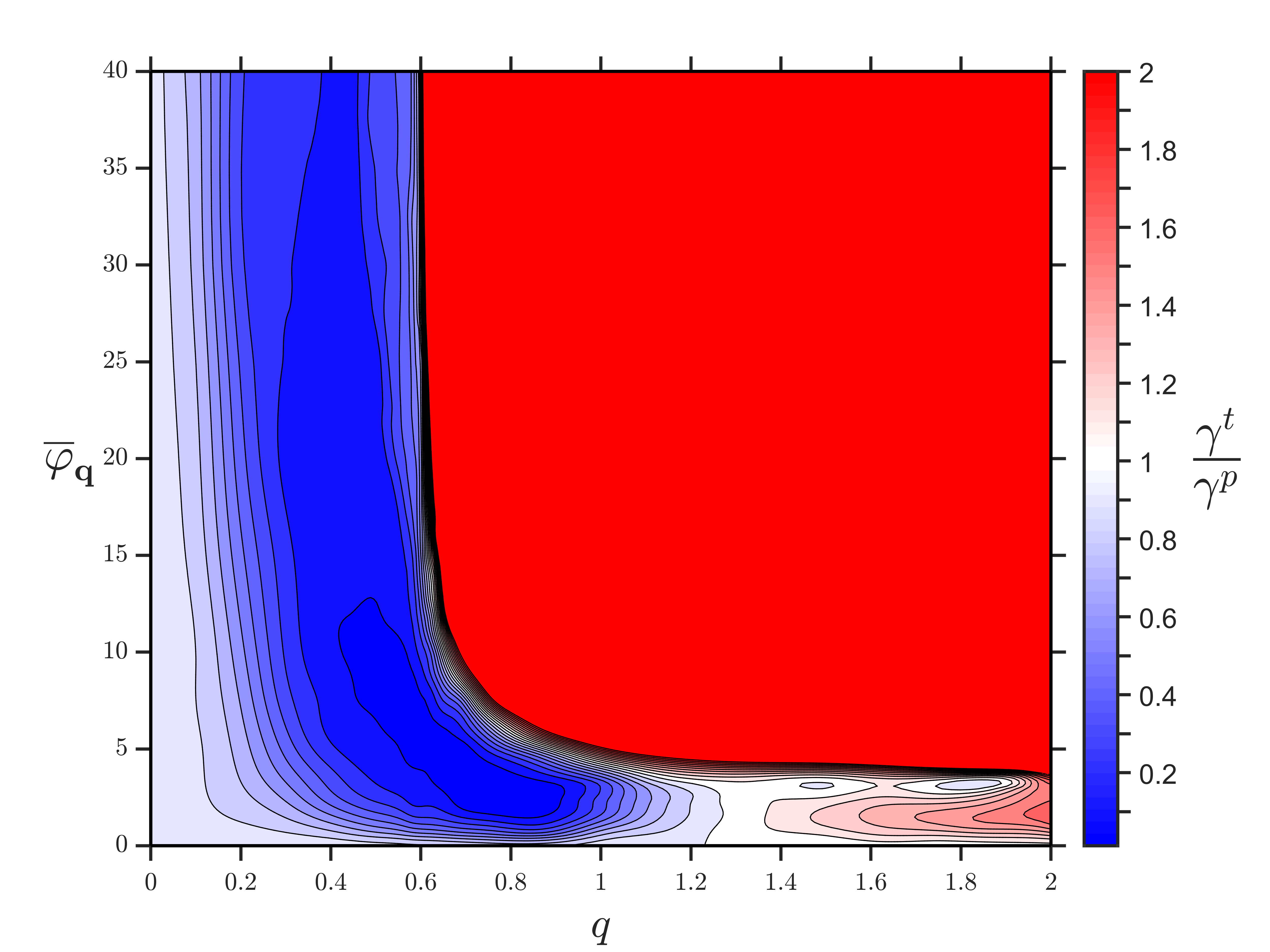}
\caption{Growth rate $\gamma^t$ at the Dimits threshold $\eta=0.25$ and $R/L_n=1.8$ of the full tertiary mode, i.e. $q_\mathrm{G}=32$ (see the convergence of figure \ref{FigureModesConvergence}), with poloidal wavenumber $p=0.6$, normalised to the primary growth rate $\gamma^p$, for a sinusoidal zonal profile $\overline{\varphi}_{\boldsymbol{k}}$ with radial wavenumber $q$ and where the ion/electron non-adiabatic responses are correctly out of phase with each other for maximum stability, like seen in figure \ref{FigureElectronIonResponse}. Strong instability commences only when $q>p$ and, unless $q$ is large, only for large zonal flows. Below this threshold the zonal flow is uniformly stabilising, to varying degrees. Note the peculiar feature that $\gamma^t$ exhibits two separate local minima with respect to the zonal amplitude $\overline{\varphi}_{\boldsymbol{k}}$ for $q\gtrsim1.3$.}
\label{FigureTertiaryKxPhi}
\end{figure}

Turning to figure \ref{FigureTertiaryKxPhi}, we can see the full (not 4M-reduced) tertiary growth rate $\gamma^t$, for the parameters chosen according to \eqref{cond5}, as a function of the zonal amplitude $\overline{\varphi}_{\boldsymbol{q}}$ and the radial wavenumber $q$. The 4M-case looks qualitatively similar, except that the stabilisation at larger zonal amplitudes is much greater, with the minimum, i.e. the solution to \eqref{cond1}-\eqref{cond3} and \eqref{cond5}, occurring for $\overline{\varphi}_{\boldsymbol{q}}$ around an order of magnitude larger than the full tertiary minimum at $q\approx0.55$ and $\overline{\varphi}_{\boldsymbol{q}}\approx5$ in figure \ref{FigureTertiaryKxPhi}. This difference hints that the direct coupling to conjugate mirror modes of \citet{Pueschel2021}, captured by the 4M-description, may easily be spoiled by the inclusion of further sidebands. The fact that typical nonlinear values, not to mention the most stable 4M-value, correspond to points far outside the valid range of a 4M-truncation, like figure \ref{FigureModesConvergence} (b) demonstrates, furthermore indicates that caution is warranted when making the reduction in only considering direct sideband coupling.

Now a stark feature is observed in figure \ref{FigureTertiaryKxPhi} for large $\overline{\varphi}_{\boldsymbol{q}}$ as $q$ crosses the boundary $q=p$. Though it is not shown, below this value there always exists \textit{stabilising} zonal distributions $g_{s\boldsymbol{q}}$ with $\gamma^t<\gamma^p$, whereas above it none cannot be found. This is the behaviour of KH-like modes \citep{Zhu2018}, with the generalised KH instability criterion that the radial wavenumber must exceed the poloidal of gyrokinetics, $k_x^2>k_y^2$, already well known \citep{Kim2002,Diamond2005}. Thus we can easily identify the unstable upper right region of \ref{FigureTertiaryKxPhi} as KH-dominated, with an associated $k_xk_y|\overline{\varphi}|$-dependent growth rate. Note however that the \textit{stabilising} profiles elsewhere nevertheless only have a finite stabilising effect and fail to actually be stable, since the value of $R/L_n$ used in figure \ref{FigureTertiaryKxPhi} corresponds to the Dimits threshold.

\subsection{Prediction comparison}\label{SectionComparison}

Finally proceeding to put the prediction into action, the $\gamma^t$-minimum of a given $R/L_n$-value, with poloidal wavenumber $p$ given by \eqref{cond1}, is obtained through a steepest-descent walk of the parameters $\varphi_{\boldsymbol{q}}$, $q$, $\alpha_1$, and $\alpha_2$ discussed in \S\;\ref{SectionAdaptation} that simultaneously solves Equations \eqref{cond2}, \eqref{cond3}, and \eqref{cond5}. Because of the presence of multiple undriven modes, the convergence of $\gamma^t$ is exceptionally slow around the Dimits threshold, and so the algorithm is commenced at $R/L_n$ well into the expected turbulent range. Having thus selected a single, minimal $\gamma^t(R/L_n)$-value, $R/L_n$ is then repeatedly lowered and a steepest descent search re-initiated to determine the slope
\begin{equation}
    \frac{\partial\gamma^t}{\partial\left(R/L_n\right)}
\end{equation}
which enables Newton's algorithm to be deployed for $\gamma^t(R/L_n)$ to find the solution to \eqref{cond4} while ameliorating the problem of slow convergence.

\begin{figure}
\centering
\includegraphics[width=0.85\linewidth]{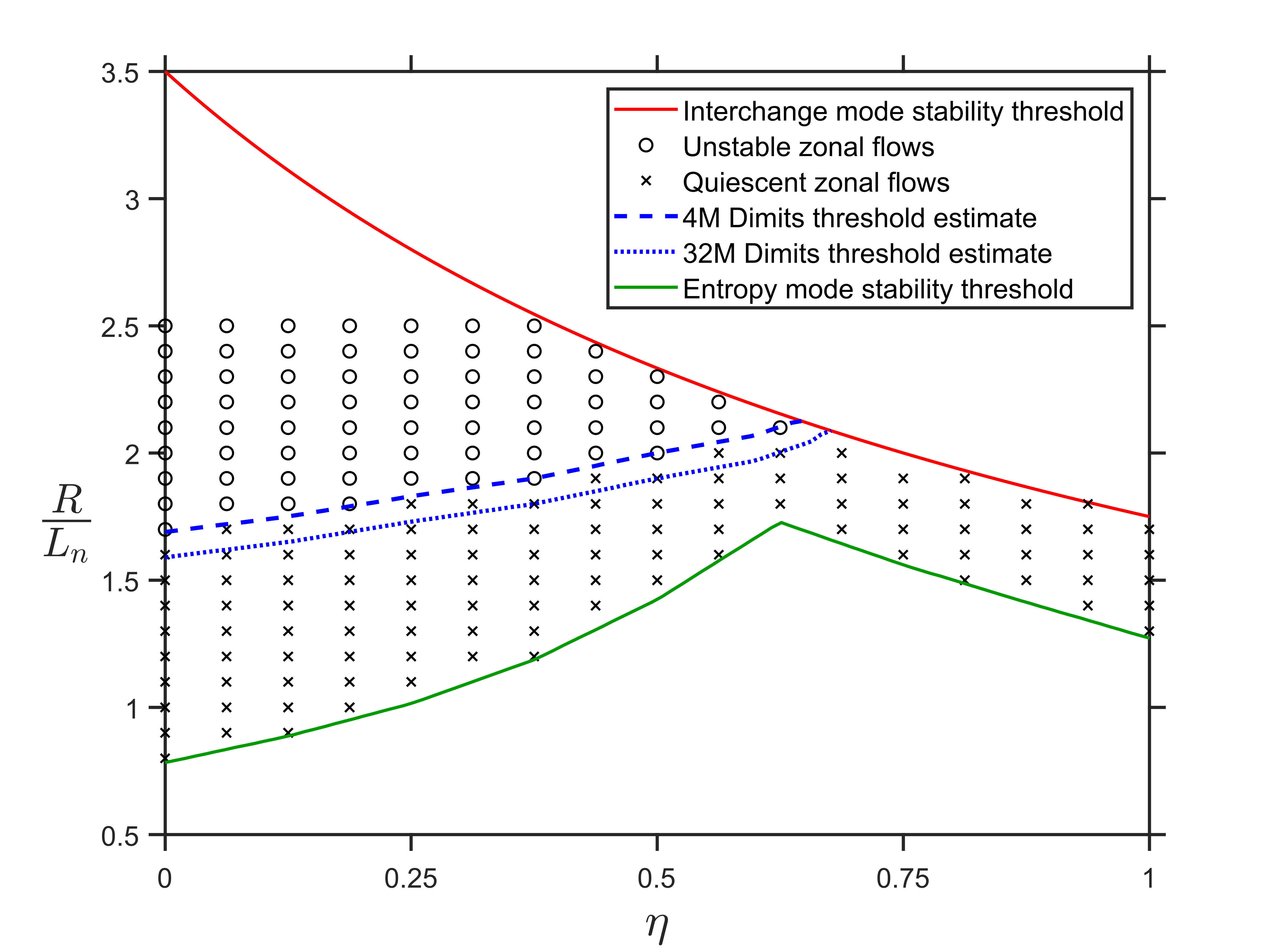}
\caption{Characterisation of the qualitative long-term behaviour of nonlinear simulations, with unstable or quiescent zonal flows, as a function of the system configuration $R/L_n$ and $\eta=L_n/L_T$. The estimated Dimits threshold, obtained as the solution to \eqref{cond1}-\eqref{cond4} and \eqref{cond5} using both a 4M and 32M calculation, is also plotted, exhibiting remarkable agreement.}
\label{FigureEndState}
\end{figure}

The result of this entire procedure compared with the result of nonlinear simulations, ran for up to $t=8000$ and categorised through the measure $\Theta_{0.1}$ of \eqref{TurbulentMeasure}, can be seen in figure \ref{FigureEndState}, both for the 4M-prediction and when effectively no restrictive truncation is employed with $q_\mathrm{G}=32$. The 4M-prediction is found to match the observed Dimits threshold remarkably well. However, as mentioned in \S\;\ref{SectionAdaptation}, the 4M solution produces high zonal amplitudes that would couple to many more sidebands if they were included. The 32M-solution does not suffer from this problem, and also does not differ much from the observed Dimits threshold. Curiously, the critical gradient of the latter case is actually lower than the former, a feature that will be discussed below. In either case, the key result is that a reduced mode tertiary instability prediction is able to accurately capture the Dimits transition.

\section{Discussion}\label{SectionDiscussion}

In this article we have performed nonlinear flux tube simulations and employed extensive associated tertiary instability analysis for the entropy-mode-driven Z-pinch using GENE. This enabled us to extend the fluid model Dimits shift prediction of \citet{Hallenbert2021}, which is essentially an efficient reduced zonal tertiary stability optimisation routine to mirror the observation that the zonal profile typically evolves towards more stable configurations. The inclusion of kinetic effects encouragingly alters the prediction minimally, and the result closely matches the observed Dimits threshold. Furthermore, the only apparent obstacle preventing this prediction from being adapted for toroidal geometries is a suitable means of accounting for the fact that only Rosenbluth-Hinton residual flows can be stable. Hopefully these can be captured by a similar trial function approach as in the present work, but more work on this front needs to be carried out. 

Compared with the gyrokinetic fluid limit of \citet{Hallenbert2021}, the Dimits regime dynamics here observed differ in some key ways, which should be kept in mind while evaluating the prediction as we go from one to the other. There, the Dimits range was characterised by large but transient turbulent bursts before the complete removal of drift waves after the emergence of a robustly stable zonal profile, i.e. one whose similar profiles are also stable so that instability is not reestablished before drift waves have completely decayed. Here, only at the very lowest Dimits range do drift waves completely decay. Instead, for a majority of the Dimits range localised, low amplitude drift waves continue to be present, but with insufficient amplitude to significantly alter the zonal profile. Correspondingly, zonal profile evolution is much slower, and the need for ``robustness" lessened, see also \S\;\ref{SectionTertiarySimulations}.

While large turbulent burst do not occur within the Dimits regime of Z-pinch gyrokinetics like those in the fluid limit, they appear as $R/L_n$ is further increased and then dominate transport \textit{before} continuous turbulence eventually develops. Indeed, these bursts seem to be of the same nature as those in \citet{Hallenbert2021}, exhibiting little change in zonal energy, rather than conventional predator-prey type zonal-drift wave oscillations \citep{Malkov2001,Kobayashi2015a}. Such oscillations arise as a result of zonal collisional damping. Higher collisionalities, though stabilising primary modes, more importantly also dampen zonal flows. This facilitates zonal/drift wave oscillations, which effectively induces higher transport and can even remove the Dimits shift entirely \citep{Kobayashi2012,Weikl2017,Ivanov2020}. Without such a mechanism here, as can be seen in figure \ref{FigureZonalEvolution} (c), instead the present bursts coincide with periods of \textit{zonal profile cycling} between profiles that are tertiary unstable at different points.

For computational efficiency, the present prediction uses a 4M reduction. Additionally, this constitutes a simplification in which only immediate sideband-coupling is considered that draws parallels to recent toroidal ITG work by \citet{Pueschel2021} and \citet{Terry2021}, who found that energy transfer rapidly diminishes in the chain of sidebands. The reason why is that direct sideband coupling can engage stable conjugate sideband mirror modes, and by a nearly zero nonlinear frequency mismatch prolonging this three-wave correlation time, this is favoured \citep{Hatch2016}.

It is helpful to contrast the view above with the conventional shearing picture, in which energy is continually shuffled, through shear, to smaller and smaller radial scale sidebands before being dissipated. Despite the difference in energy transfer, both are consistent with our tertiary picture, with poloidal bands exhibiting collective, coherent behaviour. Nevertheless, \citet{Terry2021}, based on findings of \citet{Whelan2019} that finite $\beta$-effects initially diminishes transport despite zonal flow shearing then becoming weakened, considers it likely that direct mirror mode coupling rather than zonal shearing is the dominant stabilising mechanism within the Dimits shift. For the collisionless Z-pinch however, we have three pieces of evidence that indicate that this is not the case. First, as seen in figure \ref{FigureLinear}, the Z-pinch possesses a very broad range of unstable sidebands even well inside the Dimits regime. Second, we noted that removing small-scale modes of an initially stable profile is sufficient to destabilise it. Finally, stabilisation via direct coupling in the Dimits range requires large amplitude zonal modes, whose tertiary growth rates were observed in figure \ref{FigureModesConvergence} to become unstable as more modes were included. 

With this information in hand it is natural to question the validity of a 4M-prediction, which ostensibly captures direct coupling. Therefore we completed the analogous 32M-prediction for comparison, which could also encapsulate zonal shear stabilisation. Since the maximally stabilising zonal profile found in each prediction scheme was different, it is perhaps surprising that this latter prediction was similar but slightly smaller than the previous, with both matching the observed Dimits transition well. Though there is no guarantee that this is true in general, it does hint that the \textit{potential} stabilising effect of zonal shear and conjugate mirror mode coupling may be comparable.

In \S\;\ref{SectionSingleMode}-\ref{SectionVelocitySpace} and \S\;\ref{SectionAdaptation} we saw that the choice of focusing on the Z-pinch, despite its simple geometry, was not uniformly advantageous for tertiary instability analysis. Due to the necessary inclusion of kinetic electrons and the complete conservation of zonal flows, rather than solely Rosenbluth-Hinton remnant states \citep{Rosenbluth1998a}, the linear tertiary instability picture is considerably more complicated. Both effects vastly increase the zonal configuration space which complicates the prediction method. Though we attempted to capture this, we can of course not guarantee that more stable zonal configurations can be found outside our limited selection of trial functions. The fact that the Dimits threshold, with its emergence of unstable zonal flows, so closely coincides with the point at which we cease to discover stable profiles nevertheless hints that such configurations are very rare.

On the topic of kinetic kinetic distributions, it has already been observed by \citet{Li2018b} that zonal flows need not have a uniformly stabilising effect even for configurations which are KH-stable. Thus, the stabilising influence of a zonal configuration can be altered significantly without altering the zonal flow profile itself. Nevertheless, the extent to which the full kinetic distribution can matter beyond its mere density/temperature moments has perhaps not been adequately appreciated before in the literature. At best, it is possible, though not certain, that zonal self-organisation of the kind observed here lessens the importance of this effect by precluding such resonant instability as demonstrated in figure \ref{FigureElectronIonResponse}.

Regarding the tertiary instability, there have been many recent advances in its understanding. From its introduction, it was long assumed to be a mere KH-like instability \citep{Rogers2000b}. Here we instead stress that, in accordance with recent understanding \citep{Zhu2021}, the tertiary instability should primarily be thought of as a modified primary instability, differing from the former in how it extracts energy from background gradients instead of zonal flows. Indeed, we expect the KH-instability affecting zonal flows not to play a significant role for the Z-pinch. As figure \ref{FigureTertiaryKxPhi} indicates, it could in principle effectively limit zonal amplitudes. However, since it only truly arises at zonal amplitudes quite a bit greater than what is typical in simulations, it seems dubious to expect this to actually occur. 

Now, though some attention has been given to features other than the tertiary instability in this article, naturally it has been the main focus. That is not to say that other nonlinear dynamics could not still be generally important for the Dimits threshold. Indeed, much attention has recently been paid to non-diffusive transport in the form of avalanche bursts and solitary travelling structures \citep[see e.g.][]{Ivanov2020,Qi2020}. Though no exhaustive search was carried out on this front, no immediately apparent signal indicating the presence of such features was observed. Allowing ourselves some liberty, we might speculate that their apparent absence is related to the fact that in the completely collisionless Z-pinch Dimits regime, drift wave amplitudes are too small for their self-interactions to enable manifestations of this kind. In conclusion, with the fuller understanding of the tertiary instability, observations generally point towards it being the dominant, sufficient contributor to the Dimits shift, at least for the collisionless Z-pinch.

\section*{Acknowledgements}

The authors would like to thank A. Ba\~nón Navarro for his direct help, which enabled GENE to be fruitfully employed to perform tertiary instability simulations, as well as P. Helander for his continuous support.

\section*{Funding}

This work has been carried out within the framework of the EUROfusion consortium and has received funding from the Euratom research and training programme 2014-2018 and 2019-2020 under grant agreement No 633053. The views and opinions expressed herein do not necessarily reflect those of the European Commission.

\section*{Declaration of Interests}

The authors report no conflict of interest.

\appendix

\section{Invariance of the single mode zonal profile tertiary instability}
\label{AppendixInvariance}

Recall from \S\;\ref{SectionTertiaryInstability} that the tertiary dispersion relation can be expressed as 
\begin{equation}\label{EquationAppendixTertiaryDispersion}
    \det\left[\mathsfbi{D}\left(\sum_{\boldsymbol{k}}\overline{g}_{s\boldsymbol{k}}\right)\right] = \det\left[\sum_{s=i,e}\left(\frac{Z_s}{n}\int d^3v\mathsfbi{J}_{0s}\mathsfbi{A}_s^{-1}\mathsfbi{B}_s\mathsfbi{J}_{0s}-(\mathsfbi{I}-\boldsymbol{\Gamma}_{0s})\right)\right]=0
\end{equation}
where
\begin{equation}
\boldsymbol{A}_s=(\lambda+i\omega_{ds\boldsymbol{p}})\mathsfbi{I}+\mathsfbi{C}_{1s}, \;\;\; \boldsymbol{B}_s=iZ_sf_{0s}(\omega_{*s\boldsymbol{p}}-\omega_{ds\boldsymbol{p}})\mathsfbi{I}+\mathsfbi{C}_{2s},
\end{equation}
and the elements of $\mathsfbi{C}_{1s}$, $\mathsfbi{C}_{2s}$, and $\mathsfbi{J}_{0s}$ are given by
\begin{equation}\label{EquationAppendixChain1}
    C_{1smn}=-pq\sum_{l=-j}^{j}lJ_{0sll}\overline{\varphi}_l\delta_{m(l+n)},
\end{equation}
\begin{equation}
    \;\;\; C_{2smn}=-pq\sum_{l=-j}^{j}l\overline{g}_{sl}\delta_{m(l+n)},
\end{equation}
\begin{equation}
    J_{0smn}=J_0(\sqrt{2(p^2+m^2q^2)m_s/m_i}w_s)\delta_{mn},
\end{equation} 
and 
\begin{equation}\label{EquationAppendixChain4}
    \Gamma_{0smn}=I_0((p^2+m^2q^2)m_s/m_i)e^{-(p^2+m^2q^2)m_s/m_i}\delta_{mn}.
\end{equation}

Restricting the zonal profile to 
the single mode $\boldsymbol{q}$ (and its conjugate $-\boldsymbol{q}$) we now set $\overline{\varphi}_l=\overline{g}_{sl}=0$ for $l\neq\pm1$. Then we can write
\begin{align}
    \mathsfbi{C}_{1s}=pq\begin{bNiceMatrix}[nullify-dots]
    0 & \overline{\varphi}_{\boldsymbol{q}}^* & 0 & \Cdots & & 0 \\
    -\overline{\varphi}_{\boldsymbol{q}} & \Ddots & \Ddots & \Ddots & & \Vdots \\
    0 & \Ddots & & & & \\
    \Vdots & \Ddots & & & & 0 \\
      & & & & & \overline{\varphi}_{\boldsymbol{q}}^* \\
    0 & \Cdots & & 0 & -\overline{\varphi}_{\boldsymbol{q}} & 0
    \end{bNiceMatrix} \;\;\; \mathrm{and} \;\;\; \mathsfbi{C}_{2s}=pq\begin{bNiceMatrix}[nullify-dots]
    0 & g_{s\boldsymbol{q}}^* & 0 & \Cdots & & 0 \\
    -g_{s\boldsymbol{q}} & \Ddots & \Ddots & \Ddots & & \Vdots \\
    0 & \Ddots & & & & \\
    \Vdots & \Ddots & & & & 0 \\
      & & & & & g_{s\boldsymbol{q}}^* \\
    0 & \Cdots & & 0 & -g_{s\boldsymbol{q}} & 0
    \end{bNiceMatrix},
\end{align}
where the real condition $\overline{g}_{-\boldsymbol{q}}=\overline{g}_{\boldsymbol{q}}^*$ was used. 

Next we introduce the transformation matrix
\begin{equation}
    \mathsfbi{T}=\begin{bNiceMatrix}[nullify-dots]
    0 & \Cdots & 0 & (-1)^{m} \\
    \Vdots & \Iddots & & 0 \\
    0 & \Iddots & \Iddots & \Vdots \\
    (-1)^{m} & 0 & \Cdots & 0
    \end{bNiceMatrix}
\end{equation}
where $m$ is the row number. It is clearly diagonal, $\mathsfbi{T}^T=\mathsfbi{T}$, but also orthonormal with determinant $-1$, so that $\mathsfbi{T}^2=\mathsfbi{I}$. As is easily checked, the $\mathsfbi{T}$-transformation replaces the elements of a matrix with their mirror like $E_{mn}\leftrightarrow(-1)^{m+n}E_{-m-n}$ (remember that the indexing runs from $-j$ to $j$). By \eqref{EquationAppendixChain1}-\eqref{EquationAppendixChain4}, we therefore have $\mathsfbi{T}\mathsfbi{C}_{1s}\mathsfbi{T}=\mathsfbi{C}_{1s}^*$, $\mathsfbi{T}\mathsfbi{C}_{2s}\mathsfbi{T}=\mathsfbi{C}_{2s}^*$, $\mathsfbi{T}\mathsfbi{J}_{0s}\mathsfbi{T}=\mathsfbi{J}_{0s}$, and $\mathsfbi{T}\boldsymbol{\Gamma}_{0s}\mathsfbi{T}=\boldsymbol{\Gamma}_{0s}$. Thus \eqref{EquationAppendixTertiaryDispersion} can equivalently be expressed as
\begin{align}
    \det\left(\mathsfbi{T}\mathsfbi{D}\mathsfbi{T}\right) &= \det\left[\sum_{s=i,e}\left(\frac{Z_s}{n}\int d^3v\mathsfbi{T}\mathsfbi{J}_{0s}\mathsfbi{T}^2\mathsfbi{A}_s^{-1}\mathsfbi{T}^2\mathsfbi{B}_s\mathsfbi{T}^2\mathsfbi{J}_{0s}\mathsfbi{T}-\mathsfbi{T}(\mathsfbi{I}-\boldsymbol{\Gamma}_{0s})\mathsfbi{T}\right)\right] \nonumber\\
    &=\det\left[\sum_{s=i,e}\left(\frac{Z_s}{n}\int d^3v\mathsfbi{J}_{0s}(\mathsfbi{T}\mathsfbi{A}_s\mathsfbi{T})^{-1}\mathsfbi{T}\mathsfbi{B}_s\mathsfbi{T}\mathsfbi{J}_{0s}-(\mathsfbi{I}-\boldsymbol{\Gamma}_{0s})\right)\right] \nonumber\\
    & =\det\left(\mathsfbi{D}(\overline{g}_{s\boldsymbol{q}}^*)\right) = 0,
\end{align}
i.e. the tertiary dispersion relation of a single mode zonal profile is unchanged under the substitution $\overline{g}_{s\boldsymbol{q}}\rightarrow\overline{g}_{s\boldsymbol{q}}^*$. Of course this substitution also modifies $\overline{\varphi}_{\boldsymbol{q}}$, but if we also include a translation of the periodic domain (which patently does not modify the tertiary instability) in our substitution like
\begin{equation}
    \overline{g}_{s\boldsymbol{q}} \rightarrow  \frac{\overline{g}_{s\boldsymbol{q}}^*\overline{\varphi}_{\boldsymbol{q}}^2}{|\overline{\varphi}_{\boldsymbol{q}}|^2},
\end{equation}
it is clear upon insertion into the quasineutrality condition \eqref{quasineutrality} that $\overline{\varphi}_{\boldsymbol{q}}$ now also remains unchanged.

\bibliographystyle{jpp}
\bibliography{library}

\begin{thebibliography}{91}
\expandafter\ifx\csname natexlab\endcsname\relax\def\natexlab#1{#1}\fi
\def\au#1{#1} \def\ed#1{#1} \def\yr#1{#1}\def\at#1{#1}\def\jt#1{\textit{#1}}
  \def\bt#1{#1}\def\bvol#1{\textbf{#1}} \def\vol#1{#1} \def\pg#1{#1}
  \def\publ#1{#1}\def\arxiv#1{#1}\def\org#1{#1}\def\st#1{\textit{#1}}

\bibitem[Abel {\em et~al.\/}(2013)Abel, Plunk, Wang, Barnes, Cowley, Dorland \&
  Schekochihin]{Abel2013}
{\sc \au{Abel, I~G}, \au{Plunk, G~G}, \au{Wang, E}, \au{Barnes, M}, \au{Cowley,
  S~C}, \au{Dorland, W} \& \au{Schekochihin, A~A}} \yr{2013}  \at{{Multiscale
  gyrokinetics for rotating tokamak plasmas: fluctuations, transport and energy
  flows}}.  \jt{Reports on Progress in Physics}  \bvol{76}~(11),  \pg{116201},
  \arxiv{arXiv: 1209.4782}.

\bibitem[{Ba{\~{n}}{\'{o}}n Navarro} {\em et~al.\/}(2016){Ba{\~{n}}{\'{o}}n
  Navarro}, Teaca \& Jenko]{BanonNavarro2016}
{\sc \au{{Ba{\~{n}}{\'{o}}n Navarro}, A}, \au{Teaca, B} \& \au{Jenko, F}}
  \yr{2016}  \at{{The anisotropic redistribution of free energy for gyrokinetic
  plasma turbulence in a Z-pinch}}.  \jt{Physics of Plasmas}  \bvol{23}~(4),
  \pg{042301}.

\bibitem[{Ba{\~{n}}{\'{o}}n Navarro} {\em et~al.\/}(2014){Ba{\~{n}}{\'{o}}n
  Navarro}, Teaca, Jenko, Hammett \& Happel]{BanonNavarro2014}
{\sc \au{{Ba{\~{n}}{\'{o}}n Navarro}, A}, \au{Teaca, B}, \au{Jenko, F},
  \au{Hammett, G~W} \& \au{Happel, T}} \yr{2014}  \at{{Applications of large
  eddy simulation methods to gyrokinetic turbulence}}.  \jt{Physics of Plasmas}
   \bvol{21}~(3),  \pg{032304}.

\bibitem[Belli {\em et~al.\/}(2008)Belli, Hammett \& Dorland]{Belli2008}
{\sc \au{Belli, E~A}, \au{Hammett, G~W} \& \au{Dorland, W}} \yr{2008}
  \at{{Effects of plasma shaping on nonlinear gyrokinetic turbulence}}.
  \jt{Phys. Plasmas}  \bvol{15},  \pg{969}.

\bibitem[Berionni \& G{\"{u}}rcan(2011)]{Berionni2011}
{\sc \au{Berionni, V} \& \au{G{\"{u}}rcan, {\"{O}}~D}} \yr{2011}  \at{{Predator
  prey oscillations in a simple cascade model of drift wave turbulence}}.
  \jt{Physics of Plasmas}  \bvol{18}~(11),  \pg{112301}.

\bibitem[Biglari {\em et~al.\/}(1990)Biglari, Diamond \& Terry]{Biglari1990a}
{\sc \au{Biglari, H}, \au{Diamond, P~H} \& \au{Terry, P~W}} \yr{1990}
  \at{{Influence of sheared poloidal rotation on edge turbulence}}.
  \jt{Physics of Fluids B: Plasma Physics}  \bvol{2}~(1),  \pg{1}.

\bibitem[Boozer(1998)]{Boozer1998}
{\sc \au{Boozer, A~H}} \yr{1998}  \at{{What is a stellarator?}}  \jt{Physics of
  Plasmas}  \bvol{5}~(5),  \pg{1647--1655}.

\bibitem[Bourdelle {\em et~al.\/}(2007)Bourdelle, Garbet, Imbeaux, Casati,
  Dubuit, Guirlet \& Parisot]{Bourdelle2007a}
{\sc \au{Bourdelle, C}, \au{Garbet, X}, \au{Imbeaux, F}, \au{Casati, A},
  \au{Dubuit, N}, \au{Guirlet, R} \& \au{Parisot, T}} \yr{2007}  \at{{A new
  gyrokinetic quasilinear transport model applied to particle transport in
  tokamak plasmas}}.  \jt{Physics of Plasmas}  \bvol{14}~(11),  \pg{112501}.

\bibitem[Candy {\em et~al.\/}(2004)Candy, Waltz \& Dorland]{Candy2004}
{\sc \au{Candy, J}, \au{Waltz, R~E} \& \au{Dorland, W}} \yr{2004}  \at{{The
  local limit of global gyrokinetic simulations}}.  \jt{Physics of Plasmas}
  \bvol{11}~(5),  \pg{L25--L28}.

\bibitem[Catto(1978)]{Catto1978}
{\sc \au{Catto, P~J}} \yr{1978}  \at{{Linearized gyro-kinetics}}.  \jt{Plasma
  Physics}  \bvol{20}~(7),  \pg{719}.

\bibitem[Chew {\em et~al.\/}(1956)Chew, Goldberger \& Low]{Chew1956}
{\sc \au{Chew, G~F}, \au{Goldberger, M~L} \& \au{Low, F~E}} \yr{1956}  \at{{The
  Boltzmann equation an d the one-fluid hydromagnetic equations in the absence
  of particle collisions}}.  \jt{Proceedings of the Royal Society of London.
  Series A. Mathematical and Physical Sciences}  \bvol{236}~(1204),
  \pg{112--118}.

\bibitem[Diamond {\em et~al.\/}(2005)Diamond, Itoh, Itoh \& Hahm]{Diamond2005}
{\sc \au{Diamond, P~H}, \au{Itoh, S-I}, \au{Itoh, K} \& \au{Hahm, T~S}}
  \yr{2005}  \at{{Zonal flows in plasma—a review}}.  \jt{Plasma Physics and
  Controlled Fusion}  \bvol{47}~(5),  \pg{R35}.

\bibitem[Diamond \& Kim(1991)]{Diamond1991}
{\sc \au{Diamond, P~H} \& \au{Kim, Y-B}} \yr{1991}  \at{{Theory of mean
  poloidal flow generation by turbulence}}.  \jt{Physics of Fluids B: Plasma
  Physics}  \bvol{3}~(7),  \pg{1626}.

\bibitem[Diamond {\em et~al.\/}(1994)Diamond, Liang, Carreras \&
  Terry]{Diamond1994}
{\sc \au{Diamond, P~H}, \au{Liang, Y-M}, \au{Carreras, B~A} \& \au{Terry, P~W}}
  \yr{1994}  \at{{Self-regulating shear flow turbulence: a paradigm for the L
  to H Transition}}.  \jt{Physical Review Letters}  \bvol{72}~(16),  \pg{2565}.

\bibitem[Dif-Pradalier {\em et~al.\/}(2010)Dif-Pradalier, Diamond, Grandgirard,
  Sarazin, Abiteboul, Garbet, Ghendrih, Strugarek, Ku \&
  Chang]{Dif-Pradalier2010}
{\sc \au{Dif-Pradalier, G}, \au{Diamond, P~H}, \au{Grandgirard, V},
  \au{Sarazin, Y}, \au{Abiteboul, J}, \au{Garbet, X}, \au{Ghendrih, Ph},
  \au{Strugarek, A}, \au{Ku, S} \& \au{Chang, C~S}} \yr{2010}  \at{{On the
  validity of the local diffusive paradigm in turbulent plasma transport}}.
  \jt{Physical Review E}  \bvol{82}~(2),  \pg{025401}.

\bibitem[Dimits {\em et~al.\/}(2000)Dimits, Bateman, Beer, Cohen, Dorland,
  Hammett, Kim, Kinsey, Kotschenreuther, Kritz, Lao, Mandrekas, Nevins, Parker,
  Redd, Shumaker, Sydora \& Weiland]{Dimits2000}
{\sc \au{Dimits, A~M}, \au{Bateman, G}, \au{Beer, M~A}, \au{Cohen, B~I},
  \au{Dorland, W}, \au{Hammett, G~W}, \au{Kim, C}, \au{Kinsey, J~E},
  \au{Kotschenreuther, M}, \au{Kritz, A~H}, \au{Lao, L~L}, \au{Mandrekas, J},
  \au{Nevins, W~M}, \au{Parker, S~E}, \au{Redd, A~J}, \au{Shumaker, D~E},
  \au{Sydora, R} \& \au{Weiland, J}} \yr{2000}  \at{{Comparisons and physics
  basis of tokamak transport models and turbulence simulations}}.  \jt{Physics
  of Plasmas}  \bvol{7}~(3),  \pg{969}.

\bibitem[Dorland \& Hammett(1993)]{Dorland1993}
{\sc \au{Dorland, W} \& \au{Hammett, G~W}} \yr{1993}  \at{{Gyrofluid turbulence
  models with kinetic effects}}.  \jt{Physics of Fluids B: Plasma Physics}
  \bvol{5}~(3),  \pg{812}.

\bibitem[Frieman(1982)]{Frieman1982}
{\sc \au{Frieman, E~A}} \yr{1982}  \at{{Nonlinear gyrokinetic equations for
  low-frequency electromagnetic waves in general plasma equilibria}}.
  \jt{Physics of Fluids}  \bvol{25}~(3),  \pg{502}.

\bibitem[Garbet {\em et~al.\/}(2021)Garbet, Panico, Varennes, Gillot,
  Dif-Pradalier, Sarazin, Grandgirard, Ghendrih \& Vermare]{Garbet2021}
{\sc \au{Garbet, X}, \au{Panico, O}, \au{Varennes, R}, \au{Gillot, C},
  \au{Dif-Pradalier, G}, \au{Sarazin, Y}, \au{Grandgirard, V}, \au{Ghendrih, P}
  \& \au{Vermare, L}} \yr{2021}  \at{{Wave trapping and E × B staircases}}.
  \jt{Physics of Plasmas}  \bvol{28}~(4),  \pg{042302}.

\bibitem[Hahm {\em et~al.\/}(1999)Hahm, Beer, Lin, Hammett, Lee \&
  Tang]{Hahm1999}
{\sc \au{Hahm, T~S}, \au{Beer, M~A}, \au{Lin, Z}, \au{Hammett, G~W}, \au{Lee,
  W~W} \& \au{Tang, W~M}} \yr{1999}  \at{{Shearing rate of time-dependent E×B
  flow}}.  \jt{Physics of Plasmas}  \bvol{6}~(3),  \pg{922--926}.

\bibitem[Hallenbert \& Plunk(2021)]{Hallenbert2021}
{\sc \au{Hallenbert, A} \& \au{Plunk, G~G}} \yr{2021}  \at{{Predicting the
  Dimits shift through reduced mode tertiary instability analysis in a strongly
  driven gyrokinetic fluid limit}}.  \jt{Journal of Plasma Physics}
  \bvol{87}~(5),  \pg{905870508}.

\bibitem[Hasegawa \& Mima(1978)]{Hasegawa1978}
{\sc \au{Hasegawa, A} \& \au{Mima, K}} \yr{1978}  \at{{Pseudo-three-dimensional
  turbulence in magnetized nonuniform plasma}}.  \jt{Physics of Fluids}
  \bvol{21}~(1),  \pg{87}.

\bibitem[Hasegawa \& Wakatani(1983)]{Hasegawa1983}
{\sc \au{Hasegawa, A} \& \au{Wakatani, M}} \yr{1983}  \at{{Plasma edge
  turbulence}}.  \jt{Physical Review Letters}  \bvol{50}~(9),  \pg{682}.

\bibitem[Hatch {\em et~al.\/}(2016)Hatch, Jenko, Navarro, Bratanov, Terry \&
  Pueschel]{Hatch2016}
{\sc \au{Hatch, D~R}, \au{Jenko, F}, \au{Navarro, A~Ba{\~{n}}{\'{o}}n},
  \au{Bratanov, V}, \au{Terry, P~W} \& \au{Pueschel, M~J}} \yr{2016}
  \at{{Linear signatures in nonlinear gyrokinetics: interpreting turbulence
  with pseudospectra}}.  \jt{New Journal of Physics}  \bvol{18}~(7),
  \pg{075018}.

\bibitem[Hatch {\em et~al.\/}(2011)Hatch, Terry, Jenko, Merz \&
  Nevins]{Hatch2011}
{\sc \au{Hatch, D~R}, \au{Terry, P~W}, \au{Jenko, F}, \au{Merz, F} \&
  \au{Nevins, W~M}} \yr{2011}  \at{{Saturation of gyrokinetic turbulence
  through damped eigenmodes}}.  \jt{Physical Review Letters}  \bvol{106}~(11),
  \pg{115003}.

\bibitem[Horton(1999)]{Horton1999}
{\sc \au{Horton, W}} \yr{1999}  \at{{Drift waves and transport}}.  \jt{Reviews
  of Modern Physics}  \bvol{71}~(3),  \pg{735--778}.

\bibitem[Ivanov {\em et~al.\/}(2020)Ivanov, Schekochihin, Dorland, Field \&
  Parra]{Ivanov2020}
{\sc \au{Ivanov, P~G}, \au{Schekochihin, A~A}, \au{Dorland, W}, \au{Field, A~R}
  \& \au{Parra, F~I}} \yr{2020}  \at{{Zonally dominated dynamics and Dimits
  threshold in curvature-driven ITG turbulence}}.  \jt{Journal of Plasma
  Physics}  \bvol{86}~(5),  \pg{855860502},  \arxiv{arXiv: 2004.04047}.

\bibitem[Jenko {\em et~al.\/}(2000)Jenko, Dorland, Kotschenreuther \&
  Rogers]{Jenko2000}
{\sc \au{Jenko, F}, \au{Dorland, W}, \au{Kotschenreuther, M} \& \au{Rogers,
  B~N}} \yr{2000}  \at{{Electron temperature gradient driven turbulence}}.
  \jt{Physics of Plasmas}  \bvol{7}~(5),  \pg{1904}.

\bibitem[Kadomtsev(1960)]{Kadomtsev1960}
{\sc \au{Kadomtsev, B~B}} \yr{1960}  \at{{Convective pinch instability}}.
  \jt{J. Exptl. Theoret. Phys. (U.S.S.R.)}  \bvol{37}~(10),  \pg{1096--1101}.

\bibitem[Kesner(2000)]{Kesner2000}
{\sc \au{Kesner, J}} \yr{2000}  \at{{Interchange modes in a collisional
  plasma}}.  \jt{Physics of Plasmas}  \bvol{7}~(10),  \pg{3837}.

\bibitem[Kim {\em et~al.\/}(2018)Kim, Min \& An]{Kim2018}
{\sc \au{Kim, C-B}, \au{Min, B} \& \au{An, C-Y}} \yr{2018}  \at{{Localization
  of the eigenmode of the drift-resistive plasma by zonal flow}}.  \jt{Physics
  of Plasmas}  \bvol{25}~(10),  \pg{102501}.

\bibitem[Kim {\em et~al.\/}(2019)Kim, Min \& An]{Kim2019}
{\sc \au{Kim, C-B}, \au{Min, B} \& \au{An, C-Y}} \yr{2019}  \at{{On the effects
  of nonuniform zonal flow in the resistive-drift plasma}}.  \jt{Plasma Physics
  and Controlled Fusion}  \bvol{61}~(3),  \pg{035002}.

\bibitem[Kim \& Diamond(2002)]{Kim2002}
{\sc \au{Kim, E-J} \& \au{Diamond, P~H}} \yr{2002}  \at{{Dynamics of zonal flow
  saturation in strong collisionless drift wave turbulence}}.  \jt{Physics of
  Plasmas}  \bvol{9}~(11),  \pg{4530}.

\bibitem[Kinsey {\em et~al.\/}(2005)Kinsey, Waltz \& Candy]{Kinsey2005}
{\sc \au{Kinsey, J~E}, \au{Waltz, R~E} \& \au{Candy, J}} \yr{2005}
  \at{{Nonlinear gyrokinetic turbulence simulations of E×B shear quenching of
  transport}}.  \jt{Physics of Plasmas}  \bvol{12}~(6),  \pg{062302}.

\bibitem[Kinsey {\em et~al.\/}(2007)Kinsey, Waltz \& Candy]{Kinsey2007}
{\sc \au{Kinsey, J~E}, \au{Waltz, R~E} \& \au{Candy, J}} \yr{2007}  \at{{The
  effect of plasma shaping on turbulent transport and E×B shear quenching in
  nonlinear gyrokinetic simulations}}.  \jt{Physics of Plasmas}
  \bvol{14}~(10),  \pg{102306}.

\bibitem[Kobayashi \& G{\"{u}}rcan(2015)]{Kobayashi2015}
{\sc \au{Kobayashi, S} \& \au{G{\"{u}}rcan, {\"{O}}~D}} \yr{2015}
  \at{{Gyrokinetic turbulence cascade via predator-prey interactions between
  different scales}}.  \jt{Physics of Plasmas}  \bvol{22}~(5),  \pg{050702}.

\bibitem[Kobayashi {\em et~al.\/}(2015)Kobayashi, G{\"{u}}rcan \&
  Diamond]{Kobayashi2015a}
{\sc \au{Kobayashi, S}, \au{G{\"{u}}rcan, {\"{O}}~D} \& \au{Diamond, P~H}}
  \yr{2015}  \at{{Direct identification of predator-prey dynamics in
  gyrokinetic simulations}}.  \jt{Physics of Plasmas}  \bvol{22}~(9),
  \pg{090702}.

\bibitem[Kobayashi \& Rogers(2012)]{Kobayashi2012}
{\sc \au{Kobayashi, S} \& \au{Rogers, B~N}} \yr{2012}  \at{{The quench rule,
  Dimits shift, and eigenmode localization by small-scale zonal flows}}.
  \jt{Physics of Plasmas}  \bvol{19}~(1),  \pg{012315}.

\bibitem[Kobayashi {\em et~al.\/}(2010)Kobayashi, Rogers \&
  Dorland]{Kobayashi2010}
{\sc \au{Kobayashi, S}, \au{Rogers, B~N} \& \au{Dorland, W}} \yr{2010}
  \at{{Particle pinch in gyrokinetic simulations of closed field-line
  systems}}.  \jt{Physical Review Letters}  \bvol{105}~(23),  \pg{235004}.

\bibitem[Kolesnikov \& Krommes(2005{\natexlab{{\em a\/}}})]{Kolesnikov2005}
{\sc \au{Kolesnikov, R~A} \& \au{Krommes, J~A}} \yr{2005{\natexlab{{\em a\/}}}}
   \at{{Bifurcation theory of the transition to collisionless
  ion-temperature-gradient-driven plasma turbulence}}.  \jt{Physics of Plasmas}
   \bvol{12}~(12),  \pg{122302}.

\bibitem[Kolesnikov \& Krommes(2005{\natexlab{{\em b\/}}})]{Kolesnikov2005a}
{\sc \au{Kolesnikov, R~A} \& \au{Krommes, J~A}} \yr{2005{\natexlab{{\em b\/}}}}
   \at{{Transition to collisionless ion-temperature-gradient-driven plasma
  turbulence: a dynamical systems approach}}.  \jt{Physical Review Letters}
  \bvol{94}~(23),  \pg{235002}.

\bibitem[Kraichnan(1967)]{Kraichnan1967}
{\sc \au{Kraichnan, R~H}} \yr{1967}  \at{{Inertial ranges in two-dimensional
  turbulence}}.  \jt{Physics of Fluids}  \bvol{10}~(7),  \pg{1417}.

\bibitem[Li \& Diamond(2018)]{Li2018b}
{\sc \au{Li, J~C} \& \au{Diamond, P~H}} \yr{2018}  \at{{Another look at zonal
  flows: Resonance, shearing, and frictionless saturation}}.  \jt{Physics of
  Plasmas}  \bvol{25}~(4),  \pg{042113}.

\bibitem[Liewer(1985)]{Liewer1985}
{\sc \au{Liewer, P~C}} \yr{1985}  \at{{Measurements of microturbulence in
  tokamaks and comparisons with theories of turbulence and anomalous
  transport}}.  \jt{Nuclear Fusion}  \bvol{25}~(5),  \pg{543}.

\bibitem[Lin(1998)]{Lin1998}
{\sc \au{Lin, Z}} \yr{1998}  \at{{Turbulent transport reduction by zonal flows:
  massively parallel simulations}}.  \jt{Science}  \bvol{281}~(5384),
  \pg{1835}.

\bibitem[Malkov {\em et~al.\/}(2001)Malkov, Diamond \& Rosenbluth]{Malkov2001}
{\sc \au{Malkov, M~A}, \au{Diamond, P~H} \& \au{Rosenbluth, M~N}} \yr{2001}
  \at{{On the nature of bursting in transport and turbulence in drift
  wave–zonal flow systems}}.  \jt{Physics of Plasmas}  \bvol{8}~(12),
  \pg{5073}.

\bibitem[Mantica {\em et~al.\/}(2009)Mantica, Strintzi, Tala, Giroud, Johnson,
  Leggate, Lerche, Loarer, Peeters, Salmi, Sharapov, {Van Eester}, de~Vries,
  Zabeo \& Zastrow]{Mantica2009}
{\sc \au{Mantica, P}, \au{Strintzi, D}, \au{Tala, T}, \au{Giroud, C},
  \au{Johnson, T}, \au{Leggate, H}, \au{Lerche, E}, \au{Loarer, T},
  \au{Peeters, A~G}, \au{Salmi, A}, \au{Sharapov, S}, \au{{Van Eester}, D},
  \au{de~Vries, P~C}, \au{Zabeo, L} \& \au{Zastrow, K.-D}} \yr{2009}
  \at{{Experimental study of the ion critical-gradient length and stiffness
  level and the impact of rotation in the JET tokamak}}.  \jt{Physical Review
  Letters}  \bvol{102}~(17),  \pg{175002}.

\bibitem[McMillan {\em et~al.\/}(2011)McMillan, Hill, Bottino, Jolliet, Vernay
  \& Villard]{McMillan2011}
{\sc \au{McMillan, B~F}, \au{Hill, P}, \au{Bottino, A}, \au{Jolliet, S},
  \au{Vernay, T} \& \au{Villard, L}} \yr{2011}  \at{{Interaction of large scale
  flow structures with gyrokinetic turbulence}}.  \jt{Physics of Plasmas}
  \bvol{18}~(11),  \pg{112503}.

\bibitem[McMillan {\em et~al.\/}(2009)McMillan, Jolliet, Tran, Villard, Bottino
  \& Angelino]{McMillan2009}
{\sc \au{McMillan, B~F}, \au{Jolliet, S}, \au{Tran, T~M}, \au{Villard, L},
  \au{Bottino, A} \& \au{Angelino, P}} \yr{2009}  \at{{Avalanchelike bursts in
  global gyrokinetic simulations}}.  \jt{Physics of Plasmas}  \bvol{16}~(2),
  \pg{022310}.

\bibitem[McMillan {\em et~al.\/}(2018)McMillan, Pringle \& Teaca]{McMillan2018}
{\sc \au{McMillan, B~F}, \au{Pringle, C C~T} \& \au{Teaca, B}} \yr{2018}
  \at{{Simple advecting structures and the edge of chaos in subcritical tokamak
  plasmas}}.  \jt{Journal of Plasma Physics}  \bvol{84}~(6),  \pg{905840611}.

\bibitem[Miki {\em et~al.\/}(2007)Miki, Kishimoto, Miyato \& Li]{Miki2007}
{\sc \au{Miki, K}, \au{Kishimoto, Y}, \au{Miyato, N} \& \au{Li, J~Q}} \yr{2007}
   \at{{Intermittent transport associated with the geodesic acoustic mode near
  the critical gradient regime}}.  \jt{Physical Review Letters}
  \bvol{99}~(14),  \pg{145003}.

\bibitem[Morel {\em et~al.\/}(2011)Morel, {Ba{\~{n}}{\'{o}}n Navarro},
  Albrecht-Marc, Carati, Merz, G{\"{o}}rler \& Jenko]{Morel2011}
{\sc \au{Morel, P}, \au{{Ba{\~{n}}{\'{o}}n Navarro}, A}, \au{Albrecht-Marc, M},
  \au{Carati, D}, \au{Merz, F}, \au{G{\"{o}}rler, T} \& \au{Jenko, F}}
  \yr{2011}  \at{{Gyrokinetic large eddy simulations}}.  \jt{Physics of
  Plasmas}  \bvol{18}~(7),  \pg{72301}.

\bibitem[Morel {\em et~al.\/}(2012)Morel, {Ba{\~{n}}{\'{o}}n Navarro},
  Albrecht-Marc, Carati, Merz, G{\"{o}}rler \& Jenko]{Morel2012}
{\sc \au{Morel, P}, \au{{Ba{\~{n}}{\'{o}}n Navarro}, A.}, \au{Albrecht-Marc,
  M}, \au{Carati, D}, \au{Merz, F}, \au{G{\"{o}}rler, T} \& \au{Jenko, F}}
  \yr{2012}  \at{{Dynamic procedure for filtered gyrokinetic simulations}}.
  \jt{Physics of Plasmas}  \bvol{19}~(1),  \pg{012311}.

\bibitem[Newcomb \& Kaufman(1961)]{Newcomb1961}
{\sc \au{Newcomb, Wi~A} \& \au{Kaufman, A~N}} \yr{1961}  \at{{Hydromagnetic
  stability of a tubular pinch}}.  \jt{Physics of Fluids}  \bvol{4}~(3),
  \pg{314}.

\bibitem[Numata {\em et~al.\/}(2007)Numata, Ball \& Dewar]{Numata2007}
{\sc \au{Numata, R}, \au{Ball, R} \& \au{Dewar, R~L}} \yr{2007}
  \at{{Bifurcation in electrostatic resistive drift wave turbulence}}.
  \jt{Phys. Plasmas}  \bvol{14},  \pg{82314}.

\bibitem[Parker {\em et~al.\/}(2004)Parker, Chen, Wan, Cohen \&
  Nevins]{Parker2004}
{\sc \au{Parker, S~E}, \au{Chen, Y}, \au{Wan, W}, \au{Cohen, B~I} \&
  \au{Nevins, W~M}} \yr{2004}  \at{{Electromagnetic gyrokinetic simulations}}.
  \jt{Physics of Plasmas}  \bvol{11}~(5),  \pg{2594--2599}.

\bibitem[Peeters {\em et~al.\/}(2016)Peeters, Rath, Buchholz, Camenen, Candy,
  Casson, Grosshauser, Hornsby, Strintzi \& Weikl]{Peeters2016}
{\sc \au{Peeters, A~G}, \au{Rath, F}, \au{Buchholz, R}, \au{Camenen, Y},
  \au{Candy, J}, \au{Casson, F~J}, \au{Grosshauser, S~R}, \au{Hornsby, W~A},
  \au{Strintzi, D} \& \au{Weikl, A}} \yr{2016}  \at{{Gradient-driven flux-tube
  simulations of ion temperature gradient turbulence close to the non-linear
  threshold}}.  \jt{Physics of Plasmas}  \bvol{23}~(8),  \pg{082517}.

\bibitem[Plunk {\em et~al.\/}(2014)Plunk, Helander, Xanthopoulos \&
  Connor]{Plunk2014a}
{\sc \au{Plunk, G~G}, \au{Helander, P}, \au{Xanthopoulos, P} \& \au{Connor,
  J~W}} \yr{2014}  \at{{Collisionless microinstabilities in stellarators. III.
  The ion-temperature-gradient mode}}.  \jt{Physics of Plasmas}  \bvol{21}~(3),
   \pg{032112}.

\bibitem[Pueschel {\em et~al.\/}(2016)Pueschel, Faber, Citrin, Hegna, Terry \&
  Hatch]{Pueschel2016}
{\sc \au{Pueschel, M~J}, \au{Faber, B~J}, \au{Citrin, J}, \au{Hegna, C~C},
  \au{Terry, P~W} \& \au{Hatch, D~R}} \yr{2016}  \at{{Stellarator turbulence:
  subdominant eigenmodes and quasilinear modeling}}.  \jt{Physical Review
  Letters}  \bvol{116}~(8),  \pg{085001}.

\bibitem[Pueschel \& Jenko(2010)]{Pueschel2010b}
{\sc \au{Pueschel, M~J} \& \au{Jenko, F}} \yr{2010}  \at{{Transport properties
  of finite-$\beta$ microturbulence}}.  \jt{Physics of Plasmas}  \bvol{17}~(6),
   \pg{062307}.

\bibitem[Pueschel {\em et~al.\/}(2021)Pueschel, Li \& Terry]{Pueschel2021}
{\sc \au{Pueschel, M~J}, \au{Li, P-Y} \& \au{Terry, P~W}} \yr{2021}
  \at{{Predicting the critical gradient of ITG turbulence in fusion plasmas}}.
  \jt{Nuclear Fusion}  \bvol{61}~(5),  \pg{054003}.

\bibitem[Qi {\em et~al.\/}(2020)Qi, Majda \& Cerfon]{Qi2020}
{\sc \au{Qi, D}, \au{Majda, A~J} \& \au{Cerfon, A~J}} \yr{2020}  \at{{Dimits
  shift, avalanche-like bursts, and solitary propagating structures in the
  two-field flux-balanced Hasegawa–Wakatani model for plasma edge
  turbulence}}.  \jt{Physics of Plasmas}  \bvol{27}~(10),  \pg{102304}.

\bibitem[Qian(1986)]{Qian1986}
{\sc \au{Qian, J}} \yr{1986}  \at{{Inverse energy cascade in two-dimensional
  turbulence}}.  \jt{Physics of Fluids}  \bvol{29}~(11),  \pg{3608}.

\bibitem[Qiu {\em et~al.\/}(2018)Qiu, Chen \& Zonga]{Qiu2018}
{\sc \au{Qiu, Z}, \au{Chen, L} \& \au{Zonga, F}} \yr{2018}  \at{{Kinetic theory
  of geodesic acoustic modes in toroidal plasmas: a brief review}}.  \jt{Plasma
  Science and Technology}  \bvol{20}~(9),  \pg{094004},  \arxiv{arXiv:
  1801.01622v1}.

\bibitem[Ricci {\em et~al.\/}(2006)Ricci, Rogers, Dorland \& Barnes]{Ricci2006}
{\sc \au{Ricci, P}, \au{Rogers, B~N}, \au{Dorland, W} \& \au{Barnes, M}}
  \yr{2006}  \at{{Gyrokinetic linear theory of the entropy mode in a Z pinch}}.
   \jt{Physics of Plasmas}  \bvol{13}~(6),  \pg{062102}.

\bibitem[Rogers {\em et~al.\/}(2000)Rogers, Dorland \&
  Kotschenreuther]{Rogers2000b}
{\sc \au{Rogers, B~N}, \au{Dorland, W} \& \au{Kotschenreuther, M}} \yr{2000}
  \at{{Generation and stability of zonal flows in ion-temperature-gradient mode
  turbulence}}.  \jt{Physical Review Letters}  \bvol{85}~(25),  \pg{5336}.

\bibitem[Rosenbluth \& Hinton(1998)]{Rosenbluth1998a}
{\sc \au{Rosenbluth, M~N} \& \au{Hinton, F~L}} \yr{1998}  \at{{Poloidal flow
  driven by ion-temperature-gradient turbulence in tokamaks}}.  \jt{Physical
  Review Letters}  \bvol{80}~(4),  \pg{724}.

\bibitem[Rosenbluth \& Longmire(1957)]{Rosenbluth1957}
{\sc \au{Rosenbluth, M.~N.} \& \au{Longmire, C.~L.}} \yr{1957}  \at{{Stability
  of plasmas confined by magnetic fields}}.  \jt{Annals of Physics}
  \bvol{1}~(2),  \pg{120--140}.

\bibitem[Ryter {\em et~al.\/}(2011)Ryter, Angioni, Giroud, Peeters, Biewer,
  Bilato, Joffrin, Johnson, Leggate, Lerche, Madison, Mantica, {Van Eester} \&
  Voitsekhovitch]{Ryter2011}
{\sc \au{Ryter, F}, \au{Angioni, C}, \au{Giroud, C}, \au{Peeters, A~G},
  \au{Biewer, T}, \au{Bilato, R}, \au{Joffrin, E}, \au{Johnson, T},
  \au{Leggate, H}, \au{Lerche, E}, \au{Madison, G}, \au{Mantica, P}, \au{{Van
  Eester}, D} \& \au{Voitsekhovitch, I}} \yr{2011}  \at{{Simultaneous analysis
  of ion and electron heat transport by power modulation in JET}}.  \jt{Nuclear
  Fusion}  \bvol{51}~(11),  \pg{113016}.

\bibitem[Shumlak {\em et~al.\/}(2009)Shumlak, Adams, Blakely, Chan, Golingo,
  Knecht, Nelson, Oberto, Sybouts \& Vogman]{Shumlak2009}
{\sc \au{Shumlak, U}, \au{Adams, C~S}, \au{Blakely, J~M}, \au{Chan, B-J},
  \au{Golingo, R~P}, \au{Knecht, S~D}, \au{Nelson, B~A}, \au{Oberto, R~J},
  \au{Sybouts, M~R} \& \au{Vogman, G~V}} \yr{2009}  \at{{Equilibrium, flow
  shear and stability measurements in the Z-pinch}}.  \jt{Nuclear Fusion}
  \bvol{49}~(7),  \pg{075039}.

\bibitem[Shumlak {\em et~al.\/}(2001)Shumlak, Golingo, Nelson \& {Den
  Hartog}]{Shumlak2001}
{\sc \au{Shumlak, U}, \au{Golingo, R~P}, \au{Nelson, B~A} \& \au{{Den Hartog},
  D~J}} \yr{2001}  \at{{Evidence of stabilisation in the Z-pinch}}.
  \jt{Physical Review Letters}  \bvol{87}~(20),  \pg{205005}.

\bibitem[Shumlak \& Hartman(1995)]{Shumlak1995}
{\sc \au{Shumlak, U} \& \au{Hartman, C~W}} \yr{1995}  \at{{Sheared flow
  stabilization of the m=1 kink mode in Z pinches}}.  \jt{Physical Review
  Letters}  \bvol{75}~(18),  \pg{3285--3288}.

\bibitem[Simakov {\em et~al.\/}(2002)Simakov, Hastie \& Catto]{Simakov2002}
{\sc \au{Simakov, A~N}, \au{Hastie, R~J} \& \au{Catto, P~J}} \yr{2002}
  \at{{Long mean-free path collisional stability of electromagnetic modes in
  axisymmetric closed magnetic field configurations}}.  \jt{Physics of Plasmas}
   \bvol{9}~(1),  \pg{201--211}.

\bibitem[St-Onge(2017)]{St-Onge2018}
{\sc \au{St-Onge, D~A}} \yr{2017}  \at{{On non-local energy transfer via zonal
  flow in the Dimits shift}}.  \jt{Journal of Plasma Physics}  \bvol{83}~(05),
  \pg{905830504},  \arxiv{arXiv: 1704.05406v3}.

\bibitem[Terry(2004)]{Terry2004}
{\sc \au{Terry, P~W}} \yr{2004}  \at{{Inverse energy transfer by near-resonant
  interactions with a damped-wave spectrum}}.  \jt{Physical Review Letters}
  \bvol{93}~(23),  \pg{235004}.

\bibitem[Terry {\em et~al.\/}(2006)Terry, Baver \& Gupta]{Terry2006}
{\sc \au{Terry, P~W}, \au{Baver, D~A} \& \au{Gupta, S}} \yr{2006}  \at{{Role of
  stable eigenmodes in saturated local plasma turbulence}}.  \jt{Physics of
  Plasmas}  \bvol{13}~(2),  \pg{022307}.

\bibitem[Terry {\em et~al.\/}(2018)Terry, Faber, Hegna, Mirnov, Pueschel \&
  Whelan]{Terry2018}
{\sc \au{Terry, P~W}, \au{Faber, B~J}, \au{Hegna, C~C}, \au{Mirnov, V~V},
  \au{Pueschel, M~J} \& \au{Whelan, G~G}} \yr{2018}  \at{{Saturation scalings
  of toroidal ion temperature gradient turbulence}}.  \jt{Physics of Plasmas}
  \bvol{25}~(1),  \pg{012308}.

\bibitem[Terry {\em et~al.\/}(2021)Terry, Li, Pueschel \& Whelan]{Terry2021}
{\sc \au{Terry, P~W}, \au{Li, P-Y}, \au{Pueschel, M~J} \& \au{Whelan, G~G}}
  \yr{2021}  \at{{Threshold heat-flux reduction by near-resonant energy
  transfer}}.  \jt{Physical Review Letters}  \bvol{126}~(2),  \pg{025004}.

\bibitem[Waltz {\em et~al.\/}(1998)Waltz, Dewar \& Garbet]{Waltz1998}
{\sc \au{Waltz, R~E}, \au{Dewar, R~L} \& \au{Garbet, X}} \yr{1998}  \at{{Theory
  and simulation of rotational shear stabilization of turbulence}}.
  \jt{Physics of Plasmas}  \bvol{5}~(5),  \pg{1784}.

\bibitem[Waltz {\em et~al.\/}(1994)Waltz, Kerbel \& Milovich]{Waltz1994}
{\sc \au{Waltz, R~E}, \au{Kerbel, G~D} \& \au{Milovich, J}} \yr{1994}
  \at{{Toroidal gyro‐Landau fluid model turbulence simulations in a nonlinear
  ballooning mode representation with radial modes}}.  \jt{Physics of Plasmas}
  \bvol{1}~(7),  \pg{2229}.

\bibitem[Ware(1962)]{Ware1962}
{\sc \au{Ware, A.}} \yr{1962}  \at{{Comparison between theory and experiment
  for the stability of the toroidal pinch discharge}}.  \jt{Nuclear Fusion
  Suppl.}  \bvol{3},  \pg{869}.

\bibitem[Weikl {\em et~al.\/}(2017)Weikl, Peeters, Rath, Grosshauser, Buchholz,
  Hornsby, Seiferling \& Strintzi]{Weikl2017}
{\sc \au{Weikl, A}, \au{Peeters, A~G}, \au{Rath, F}, \au{Grosshauser, S~R},
  \au{Buchholz, R}, \au{Hornsby, W~A}, \au{Seiferling, F} \& \au{Strintzi, D}}
  \yr{2017}  \at{{Ion temperature gradient turbulence close to the finite heat
  flux threshold}}.  \jt{Physics of Plasmas}  \bvol{24}~(10),  \pg{102317}.

\bibitem[Whelan {\em et~al.\/}(2019)Whelan, Pueschel, Terry, Citrin, McKinney,
  Guttenfelder \& Doerk]{Whelan2019}
{\sc \au{Whelan, G~G}, \au{Pueschel, M~J}, \au{Terry, P~W}, \au{Citrin, J},
  \au{McKinney, I~J}, \au{Guttenfelder, W} \& \au{Doerk, H}} \yr{2019}
  \at{{Saturation and nonlinear electromagnetic stabilization of ITG
  turbulence}}.  \jt{Physics of Plasmas}  \bvol{26}~(8),  \pg{082302}.

\bibitem[Winsor {\em et~al.\/}(1968)Winsor, Johnson \& Dawson]{Winsor1968}
{\sc \au{Winsor, N}, \au{Johnson, J~L} \& \au{Dawson, J~M}} \yr{1968}
  \at{{Geodesic acoustic waves in hydromagnetic systems}}.  \jt{Physics of
  Fluids}  \bvol{11}~(11),  \pg{2448}.

\bibitem[van Wyk {\em et~al.\/}(2016)van Wyk, Highcock, Schekochihin, Roach,
  Field \& Dorland]{VanWyk2016}
{\sc \au{van Wyk, F}, \au{Highcock, E~G}, \au{Schekochihin, A~A}, \au{Roach,
  C~M}, \au{Field, A~R} \& \au{Dorland, W}} \yr{2016}  \at{{Transition to
  subcritical turbulence in a tokamak plasma}}.  \jt{Journal of Plasma Physics}
   \bvol{82}~(6),  \pg{905820609},  \arxiv{arXiv: 1607.08173}.

\bibitem[Xanthopoulos {\em et~al.\/}(2007)Xanthopoulos, Merz, G{\"{o}}rler \&
  Jenko]{Xanthopoulos2007}
{\sc \au{Xanthopoulos, P}, \au{Merz, F}, \au{G{\"{o}}rler, T} \& \au{Jenko, F}}
  \yr{2007}  \at{{Nonlinear gyrokinetic simulations of ion-temperature-gradient
  turbulence for the optimized Wendelstein 7-X stellarator}}.  \jt{Physical
  Review Letters}  \bvol{99}~(3),  \pg{035002}.

\bibitem[Zhang \& Krasheninnikov(2020)]{Zhang2020}
{\sc \au{Zhang, Y} \& \au{Krasheninnikov, S~I}} \yr{2020}  \at{{Influence of
  zonal flow and density on resistive drift wave turbulent transport}}.
  \jt{Physics of Plasmas}  \bvol{27}~(12),  \pg{122303}.

\bibitem[Zhu \& Dodin(2021)]{Zhu2021}
{\sc \au{Zhu, H} \& \au{Dodin, I~Y}} \yr{2021}  \at{{Wave-kinetic approach to
  zonal-flow dynamics: Recent advances}}.  \jt{Physics of Plasmas}
  \bvol{28}~(3),  \pg{032303}.

\bibitem[Zhu {\em et~al.\/}(2018)Zhu, Zhou \& Dodin]{Zhu2018}
{\sc \au{Zhu, H}, \au{Zhou, Y} \& \au{Dodin, I~Y}} \yr{2018}  \at{{On the
  Rayleigh–Kuo criterion for the tertiary instability of zonal flows}}.
  \jt{Physics of Plasmas}  \bvol{25}~(8),  \pg{082121},  \arxiv{arXiv:
  1805.02233}.

\bibitem[Zhu {\em et~al.\/}(2020{\natexlab{{\em a\/}}})Zhu, Zhou \&
  Dodin]{Zhu2020}
{\sc \au{Zhu, H}, \au{Zhou, Y} \& \au{Dodin, I~Y}} \yr{2020{\natexlab{{\em
  a\/}}}}  \at{{Theory of the tertiary instability and the Dimits shift from
  reduced drift-wave models}}.  \jt{Physical Review Letters}  \bvol{124}~(5),
  \pg{055002}.

\bibitem[Zhu {\em et~al.\/}(2020{\natexlab{{\em b\/}}})Zhu, Zhou \&
  Dodin]{Zhu2020a}
{\sc \au{Zhu, H}, \au{Zhou, Y} \& \au{Dodin, I~Y}} \yr{2020{\natexlab{{\em
  b\/}}}}  \at{{Theory of the tertiary instability and the Dimits shift within
  a scalar model}}.  \jt{Journal of Plasma Physics}  \bvol{86}~(4),
  \pg{905860405}.

\end{thebibliography}

\end{document}